\documentclass[sigconf, nonacm]{acmart}
\AtBeginDocument{%
  \providecommand\BibTeX{{%
    \normalfont B\kern-0.5em{\scshape i\kern-0.25em b}\kern-0.8em\TeX}}}

\authorsaddresses{}              
\renewcommand{\shortauthors}{}

\setcopyright{acmlicensed}
\copyrightyear{2018}
\acmYear{2018}
\acmDOI{XXXXXXX.XXXXXXX}

\acmJournal{JDS}
\acmVolume{37}
\acmNumber{4}
\acmArticle{111}
\acmMonth{8}

\usepackage{microtype}
\usepackage{graphicx}
\usepackage{subcaption}
\usepackage{booktabs} 
\usepackage{pifont}
\usepackage{caption}
\usepackage{amsmath}
\usepackage{dsfont}
\usepackage{circledsteps}
\usepackage{lipsum}
\usepackage{stfloats} 
\usepackage{enumitem}
\usepackage{hyperref}
\usepackage{algorithm}
\usepackage{algpseudocode}
\usepackage{hyperref}
\usepackage{multirow}

\DeclareMathOperator*{\argmax}{arg\,max}

\newcommand{\name}{FLAMMABLE}

\newcommand{\mmfl}{Multi-Model Federated Learning}

\newcommand{\modify}[1]{{ #1}}
\newcommand{\specialcell}[2][c]{%
  \begin{tabular}[#1]{@{}c@{}}#2\end{tabular}}

\begin{document}

\title{The Name of the Title Is Hope}

\author{Shouxu Lin}
\affiliation{%
  \institution{Cornell University}
  \city{Ithaca}
  \state{NY}
  \country{USA}
}

\author{Zimeng Pan}
\affiliation{%
  \institution{Carnegie Mellon University}
  \city{Pittsburgh}
  \state{PA}
  \country{USA}}

\author{Yuhang Yao}
\affiliation{%
  \institution{Carnegie Mellon University}
  \city{Pittsburgh}
  \state{PA}
  \country{USA}}

\author{Haeyoung Noh}
\affiliation{%
  \institution{Stanford University}
  \city{Palo Alto}
  \state{CA}
  \country{USA}}

\author{Pei Zhang}
\affiliation{%
  \institution{University of Michigan}
  \city{Ann Arbor}
  \state{MI}
  \country{USA}}

\author{Carlee Joe-Wong}
\affiliation{%
  \institution{Carnegie Mellon University}
  \city{Pittsburgh}
  \state{PA}
  \country{USA}}


\renewcommand{\shortauthors}{}

\acmArticleType{Review}
\acmCodeLink{https://github.com/borisveytsman/acmart}
\acmDataLink{htps://zenodo.org/link}
\acmContributions{BT and GKMT designed the study; LT, VB, and AP
  conducted the experiments, BR, HC, CP and JS analyzed the results,
  JPK developed analytical predictions, all authors participated in
  writing the manuscript.}

\title[\name{}]{\name{}: A Multi-Model Federated Learning Framework with Multi-Model Engagement and Adaptive Batch Sizes}

\begin{abstract}
Multi-Model Federated Learning (MMFL) is an emerging direction in Federated Learning (FL) where multiple models are trained in parallel, generally on various datasets.  Optimizing the models' accuracies and training times in the MMFL setting requires adapting to data and system heterogeneity across clients as in single-model FL; these challenges are amplified in the MMFL setting due to additional heterogeneity across models. Neither existing solutions nor na\"ive extensions of single-model FL frameworks efficiently address these challenges. To bridge this gap, we propose \name{}, a comprehensive MMFL training framework.
\name{} optimizes model training by intelligently adapting client batch sizes while engaging them to train multiple carefully chosen models, depending on their system capabilities, in each training round. To evaluate \name{}, we develop the first benchmark platform for the MMFL setting, which may enable future reproducible MMFL research. Extensive evaluations on multiple datasets and models show that \name{} boosts the MMFL time-to-accuracy performance by 1.1$\sim$10.0$\times$ while improving the final model accuracy by 1.3$\sim$5.4\% compared to several known baselines.
\end{abstract}

\maketitle

\section{Introduction}

\begin{figure}[t]
    \centering
    \includegraphics[trim=0 80pt 0 20pt, clip, width=1\linewidth]{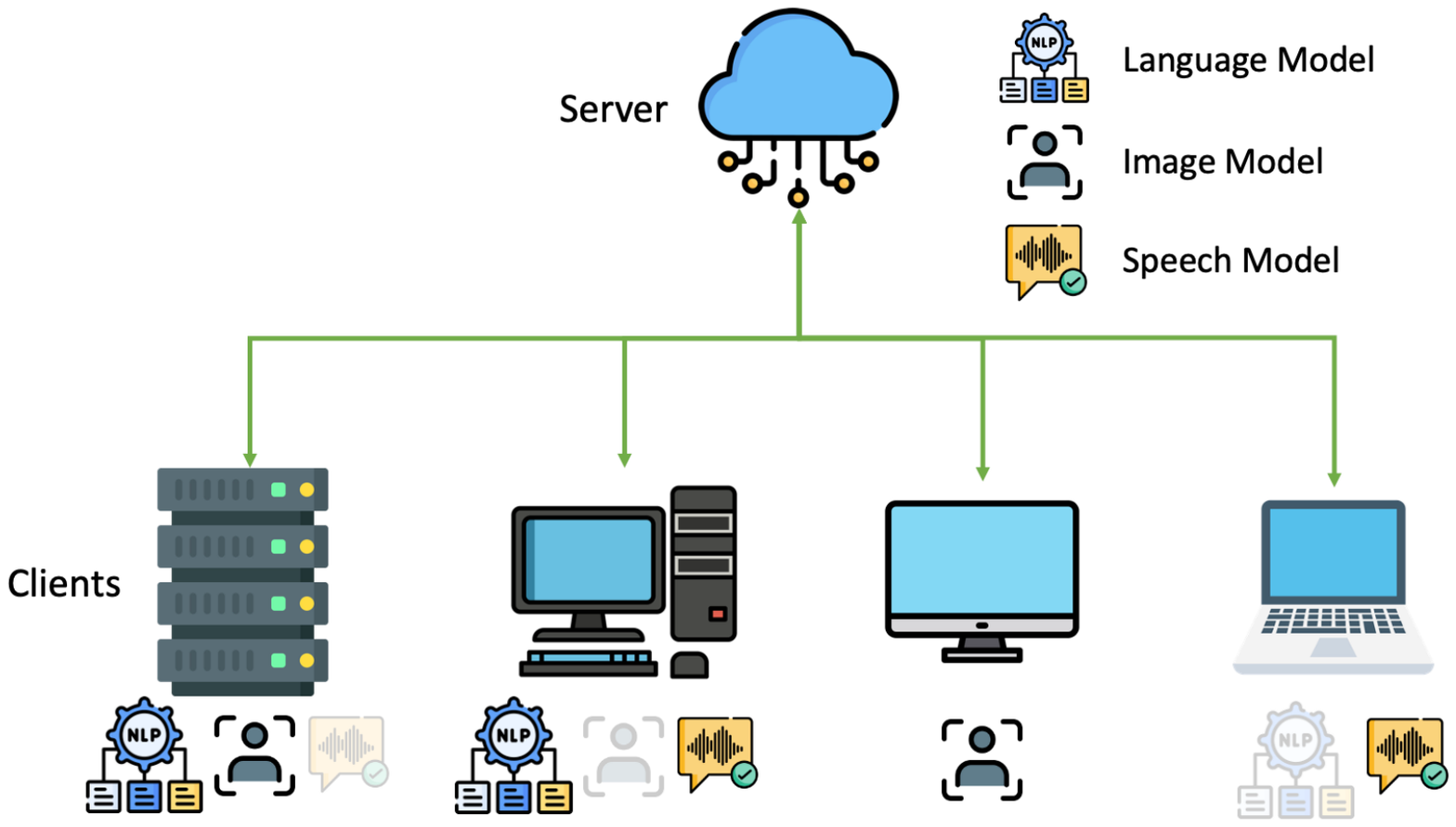}
    \caption{Overview of Multi-Model Federated Learning. Each client has the local data for training a subset of models. During every training round, the central server decides which client should train which model(s), shown by the darker-colored models in the figure. A faster client can train another model in the same round once it has finished its training tasks for other models.}
    \label{fig:intro_image}
\end{figure}

Federated Learning (FL) is an innovative machine learning paradigm that enables model training across numerous clients holding local data~\cite{fedavg, wang2020federated}. FL amounts to producing a generalized representation for a distributed dataset against a specified model architecture. Initially suggested for next-word prediction in the context of data protection legislations like GDPR (General Data Protection Regulation) in Europe, the FL approach has proven to be versatile and has been extended to various types of data and model architecture~\cite{li2020review, li2020federated, aledhari2020federated}. For instance, individual efforts have been made in the past to analyze time-series mobile health data for health trend prediction~\cite{xu2021federated}, location data for location-based services~\cite{kong2021privacy}, and shopping data to develop recommendation systems based on past purchases~\cite{zhang2021survey}. Due to the privacy-sensitive nature of such user data, FL has even become a popular method to finetune foundation models, customizing them to private user data~\cite{zhuang2023foundation,chen2024feddat}. It is likely to soon anticipate demands for training all proposed tasks together in parallel over the same group of clients, with each client participating in one or multiple model training depending on their local data. As a result, there exists a strong need for parallel learning of multiple models in a federated manner.

A promising approach for enabling parallel FL is \textit{Multi-Model Federated Learning} (MMFL), as shown in Figure~\ref{fig:intro_image}. Similar to single-model FL, MMFL follows the FL process: a coordinator server in each round selects a subset of clients for each model to perform local model updates based on their local data, and then aggregates the local updates to obtain a global model before advancing to the next round~\cite{siew2023fair,MM_ucb,round_robin, eds_multi_aaai, logfair}.
Individual models may stop training if their target accuracy is reached, while others continue~\cite{askin2024fedast}.
Foundational works in MMFL have emphasized the importance of optimizing time-to-accuracy performance (the wall clock time for each model to achieve a target accuracy). However, previous MMFL solutions rely on assumptions that can lead to inefficient time-to-accuracy outcomes. We outline the limitations of these existing works below.

\textbf{Limitation 1: Constant batch sizes and number of local iterations.}
The statistical progress (e.g., increment of validation accuracy) of each selected client in each round must be balanced so that the aggregated global model will not be biased to some clients' data. 
To ensure this, existing works~\cite{oort, eds_multi_aaai, fedbalancer, round_robin, logfair, fedavg} simply assume that all the clients consistently perform the same number of iterations $k$ (the number of training passes over its local data) with the same batch size $m$ (the number of training samples in each pass) throughout the entire training process.
Note that the training time of a client is the product of the reciprocal of the system throughput $\theta$ (number of training samples processed per second) and the total number of samples $m \times k$.
Previous studies~\cite{wang2019benchmarking, liu2019performance} indicate that increasing batch sizes over the course of the training up to a point can improve system throughput as the hardware can process samples in parallel, thereby reducing the time to train the same number of samples.
However, the batch size maximizing system throughput varies across clients due to heterogeneous computing and memory resources. This variation grows in MMFL due to diverse model complexities and sample sizes.
However, existing works cannot optimize the batch size of each client-model pair and usually set a constant batch size agnostic to both clients and models (often small to accommodate clients with limited capacity), thus offering more opportunities to choose the wrong batch size and leading to sub-optimal client training time and time-to-accuracy performance.

\textbf{Limitation 2: Single-model engagement.}
FL usually operates in a synchronous manner~\cite{wang2019adaptive}, which offers better convergence and model quality~\cite{ wang2019adaptive, chen2016revisiting, chen2019round}. Thus, faster clients stay idle when awaiting slower clients, which may slow overall time-to accuracy. MMFL exacerbates idle times due to disparities in execution times among models, as clients training smaller models must wait for resource-demanding models to complete. However, this setting also suggests a potential solution of \textit{multi-model engagement}: assigning fast clients with multiple models. Existing MMFL solutions~\cite{eds_multi_aaai,round_robin} assume that each client can only train one model per round. 

In this paper, we introduce \textbf{\name{}} (Federated Learning Adopted for Multiple Models with Adaptive Batches for Learning Efficiently), a comprehensive MMFL framework that automatically adapts the batches (both batch size and number of iterations) tailored for each client and model, and optimizes the engagement of multiple models to each client.
To realize \name{}, we address the following challenges.

\begin{figure}[t]
    \centering
    \includegraphics[width=1\linewidth]{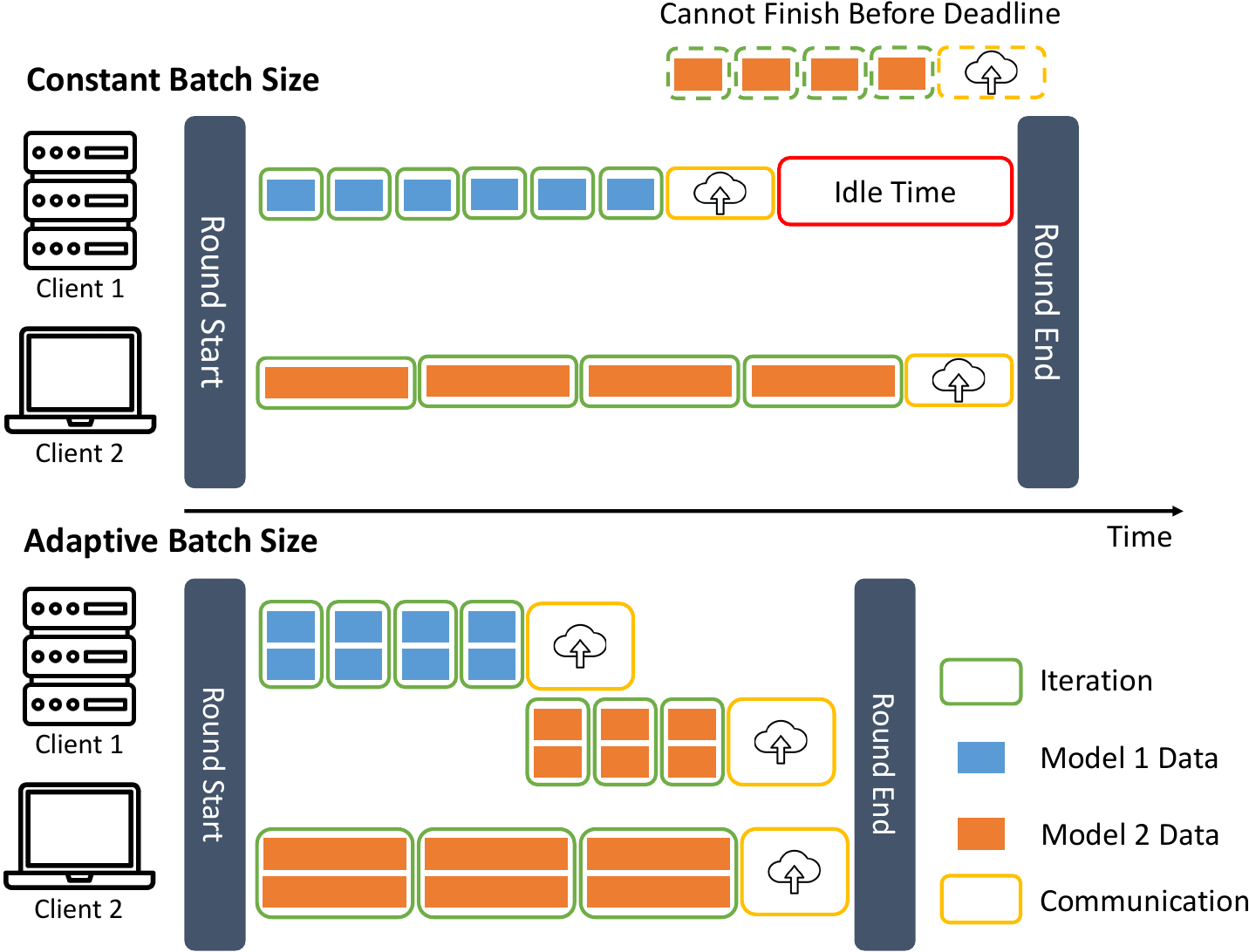}
    \caption{\name{} (1) reduces round duration by adapting the batch sizes (denoted by the height of each iteration): powerful devices can process an iteration with larger batch size with no/little increase in the time to process one iteration, (2) adapts the number of iterations per client to maintain training progress after batch size adaptation due to reduced statistical efficiency, and (3) allocates multiple models to fast clients to reduce idle time.}
    \label{fig:font_page}
\end{figure}

\textbf{Challenge 1: How to adapt batches to speedup training time without compromising statistical progress?}
A straightforward approach to batch size adaptation involves selecting the batch size with the highest throughput for each client, while maintaining the same number of data samples used for training in each round, e.g., halving the number of iterations when doubling the batch size. However, using an excessively large batch size from the onset of the training can negatively impact the final model accuracy~\cite{fedavg}. Increasing the batch size also reduces \textit{statistical efficiency}, i.e., the per-sample training contribution~\cite{johnson2019adascale}. Therefore, for clients with larger batch sizes, each training sample contributes less to learning~\cite{mccandlish2018empirical}, culminating in less overall \textit{statistical progress} (statistical efficiency $\times$ number of samples) in each round.
A solution is to co-adapt the number of iterations and the batch size to increase the total number of samples used for training. For example, as depicted in Figure~\ref{fig:font_page}, matching the progress of training 6 samples over 6 iterations with a batch size of 1 requires training 8 samples over 4 iterations with a batch size of 2. However, precisely estimating the requisite increase in the number of training samples needed after batch size adaptation is difficult. Worse, this required increment varies both across models and over the course of the training~\cite{johnson2019adascale}.
In addition, the batch size with the highest system throughput might not optimize the training time (number of samples $\div$ system throughput) to achieve the same statistical efficiency due to the increase in the number of training samples required.

\name{} first quantifies the relationship between statistical efficiency and batch sizes and accurately predicts the additional number of samples for different batch sizes required to achieve the same statistical progress w.r.t (with respect to) the default batch size and number of iterations. To do so, \name{} uses the \modify{Gradient Noise Scale captured in previous training rounds~\cite{johnson2019adascale, mccandlish2018empirical}, and then picks the batch size that minimizes training time.}

\textbf{Challenge 2: How to allocate clients to models and integrate multi-model engagement in client selection?} 
Different clients exhibit heterogeneous system behavior (e.g., training time) and data quality (e.g., the suitability and effectiveness of their data for contributing to the global model). The presence of multiple models further complicates this challenge by adding another dimension of heterogeneity, as different models may themselves have different training times and data quality at different clients. \name{} evaluates each client's utility by considering both system behavior and data quality, aiming to allocate clients to models in a way that maximizes the combined utility across all models.

A na\"ive way of multi-model client engagement is to first select clients with the highest utility for training and then allocate more models to fast clients. Decoupling these two steps, however, can lead to suboptimal client selection. Consider a scenario where a client offers greater utility for model 1 compared to models 2 and 3. If we initially allocate model 1 to this client, then attempt to allocate additional models in a subsequent step, we might find the client unable to complete any more models before the round deadline. Conversely, if this client had initially been allocated models 2 and 3, they could have completed both within the deadline, resulting in a cumulative benefit surpassing that of model 1 alone. 
\name{} thus aims to embed multi-model client allocation directly within the initial client selection process by formulating the client selection problem as a Multiple Knapsacks Problem and then transferring it to an ILP (Interger Linear Programming) through auxiliary variables which can be rapidly solved.


We summarize our \textbf{research contributions} below.
\begin{itemize}

    \item  We proposed the \textbf{first FL batch adaptation mechanism} (\S\ref{sec:system_adapt}), which automatically adapts and optimizes the batch sizes and number of iterations for each client-model pair at different stages of the training process, while matching training progress w.r.t. the progress with default batch sizes and number of iterations. To do so, we quantify the additional number of training samples and iterations needed after batch size adaptation.

    \item We designed the \textbf{first client selection strategy} (\S\ref{sec:selection}) to optimize client-to-model allocation by jointly considering client utilities across models and the viability of multi-model engagement in each round.

     \item We built the \textbf{first multi-model FL benchmark platform} (\S\ref{sec:impl}) to simply and standardize the evaluation of MMFL in various settings, as existing platforms~\cite{lai2022fedscale,he2020fedml} can only support a single model. The platform provides a backend to parallelize training on GPUs/CPUs in the cluster and an automated batch size adaption mechanism so that users can try different approaches on top of adaptive batch sizes. 

    \item  We implemented \name{} and six popular client selection frameworks used in today’s FL deployments in our platform, and \textbf{evaluated} them across various FL tasks with real-world datasets (\S\ref{sec:end_to_end_perf}). Compared to these works, \name{} improves the time-to-accuracy by 1.1$\sim$10.0$\times$ while improving the final accuracy by 1.3$\sim$5.4\%.

\end{itemize}

The rest of the paper is organized as follows. We discuss the related work in $\S\ref{sec:related_work}$.
We present the motivation study of batch adaptation and multi-model engagement in $\S\ref{sec:motivation_study}$.
Then, we discuss the overview of \name{} in \S\ref{sec:system_overview} and explain our methods of batch adaptation and client selection with multi-model engagement in \S\ref{sec:system_design}.
The experimental results are given in $\S\ref{sec:evaluation}$.
Finally, $\S\ref{sec:conclusion}$ concludes the paper and discusses future directions.

\section{Related Work}
\label{sec:related_work}

\textbf{Batch adaptation.} Most existing FL studies~\cite{fedavg, oort, fedbalancer, eds_multi_aaai, logfair, round_robin}, simply set a constant batch size and number of iterations for all the clients throughout the entire training process.
Although several works~\cite{adapt_ample, wang2019adaptive,liu2023adacoopt,ma2023adaptive} investigate the potential of adapting batch sizes, our approach presents a dual advantage.

Firstly, \textit{\name{} is the first to explore batch adaptation while matching training progress w.r.t the progress with default batch sizes and number of iterations.}
Existing works adapt batch sizes with the aim to mitigate straggler effects by harmonizing different clients' round execution times through tailored batch sizes, particularly enlarging batch sizes for faster clients~\cite{liu2023adacoopt,adapt_ample}. However, these works do not adapt the number of iterations accordingly after adapting the batch sizes. In this way, clients with larger batch sizes will make more statistical progress as they train the model with more number of samples. Thus, the aggregated global model might be biased to these clients' local data~\cite{cho2022towards, abay2020mitigating}, leading to degradation in final model quality especially when data is not identically and independently distributed (non-IID) across different clients~\cite{zhu2021federated, zhao2018federated} (to be discussed in \S\ref{fig:motivation_bsz}).
To solve this problem, \name{} quantifies the relationship between statistical efficiency and batch sizes and accurately predicts the number of samples for different batch sizes required to achieve the
same statistical progress. \name{} thus co-adapts the number of iterations after batch size adaptation to match the same statistical progress.

In addition, \textit{\name{} is the first to explore batch adaptation to expedite training progress and thus reduce time-to-accuracy performance.}
The batch sizes of slower clients remain unadapted in existing works, thereby failing to reduce the round duration, which continues to be bottlenecked by the slowest client. However, \name{} picks the batch size that can minimize the training time to achieve the same statistical progress.

Other studies, e.g., ~\cite{pollux}, optimize the batch size for centralized training, with all training samples at a single client. 
However, FL requires aggregating updates from multiple clients. Disparate progress across clients can slow convergence, making it crucial to concurrently adjust batch sizes and iteration numbers to maintain consistent statistical progress across heterogeneous clients. To the best of our knowledge our work is the \textit{first to explore batch size adaptation in FL to optimize client system throughput.}


\textbf{Existing client selection methods.}
To the best of our knowledge, our work is the \textit{first to exploit clients' system heterogeneity so that some FL clients can train multiple models in a single training round}, reducing client idle time and improving time-to-accuracy performance.
Extending single-model client selection works~\cite{oort,fedbalancer} in the \mmfl{} (MMFL) setting generates suboptimal solutions, as they cannot evaluate and compare clients' performance across different models. Prior MMFL work handles this issue by evaluating and optimizing the clients selected to train each model in each training round, under the assumption that each client can only train one model per round. These prior works propose a variety of schemes under this assumption, including round-robin~\cite{round_robin}, multi-armed bandit~\cite{MM_ucb}, or Bayesian optimization-based~\cite{eds_multi_aaai} selection algorithms. These typically seek to ensure that each model is trained on clients with high system throughput or high data quality, thus ensuring the eventual accuracy of the model, while also seeking to minimizing the overall model training time. Other work has proposed optimizing client selection in asynchronous MMFL training algorithms~\cite{chang2023asynchronous,askin2024fedast}, which similarly assumes that clients can train only one model at a time. \modify{Moreover, none of these works concurrently optimize the batch sizes.}

\modify{
Some \mmfl{} approaches~\cite{smith2017federated, zhu2024fedtrans, chen2021matching} have been proposed to handle multiple tasks. 
The major difference between the multi-task job and multi-model job is that the tasks of the multi-task jobs share some common parts of the model, while the multi-model jobs in our work do not have interaction between each other in terms of the model. The multi-model FL deals with the simultaneous training process of multiple \textit{independent} jobs.}

\section{Motivation Study}
\label{sec:motivation_study}

This section explores the underlying motivations driving our research, demonstrating that (i) na\"ively adapting the batch size can slow the time-to-accuracy ($\S\ref{sec:motivation_adapt}$), motivating the intelligent batch adaptation methods that we design in $\S\ref{sec:system_adapt}$; and (ii) engaging more clients in a given training round, as in multi-model engagement, can significantly increase time-to-accuracy ($\S\ref{sec:motivation_multi_task}$), motivating the client selection methods we design in $\S\ref{sec:selection}$.


\subsection{Batch Size Adaptation}
~\label{sec:motivation_adapt}

Existing FL works 
aim to shorten the duration of each training round by either reducing the amount of training data used~\cite{fedbalancer} or selecting clients with superior system throughput~\cite{oort,eds_multi_aaai}. The first approach is only effective within the epoch training framework, where each client trains over the entire local dataset. Most FL deployments, however, use the iteration training framework, where clients only need to train over a subset of local data in a few number of iterations. Furthermore, while reducing the amount of data used can reduce the per-round training time~\cite{ruan2021towards}, it does not improve the overall throughput. 
The second approach, though effective at prioritizing system throughput, fails to further optimize the chosen clients' throughput.

\begin{figure}[h]
    \centering
    \includegraphics[width=1\linewidth]{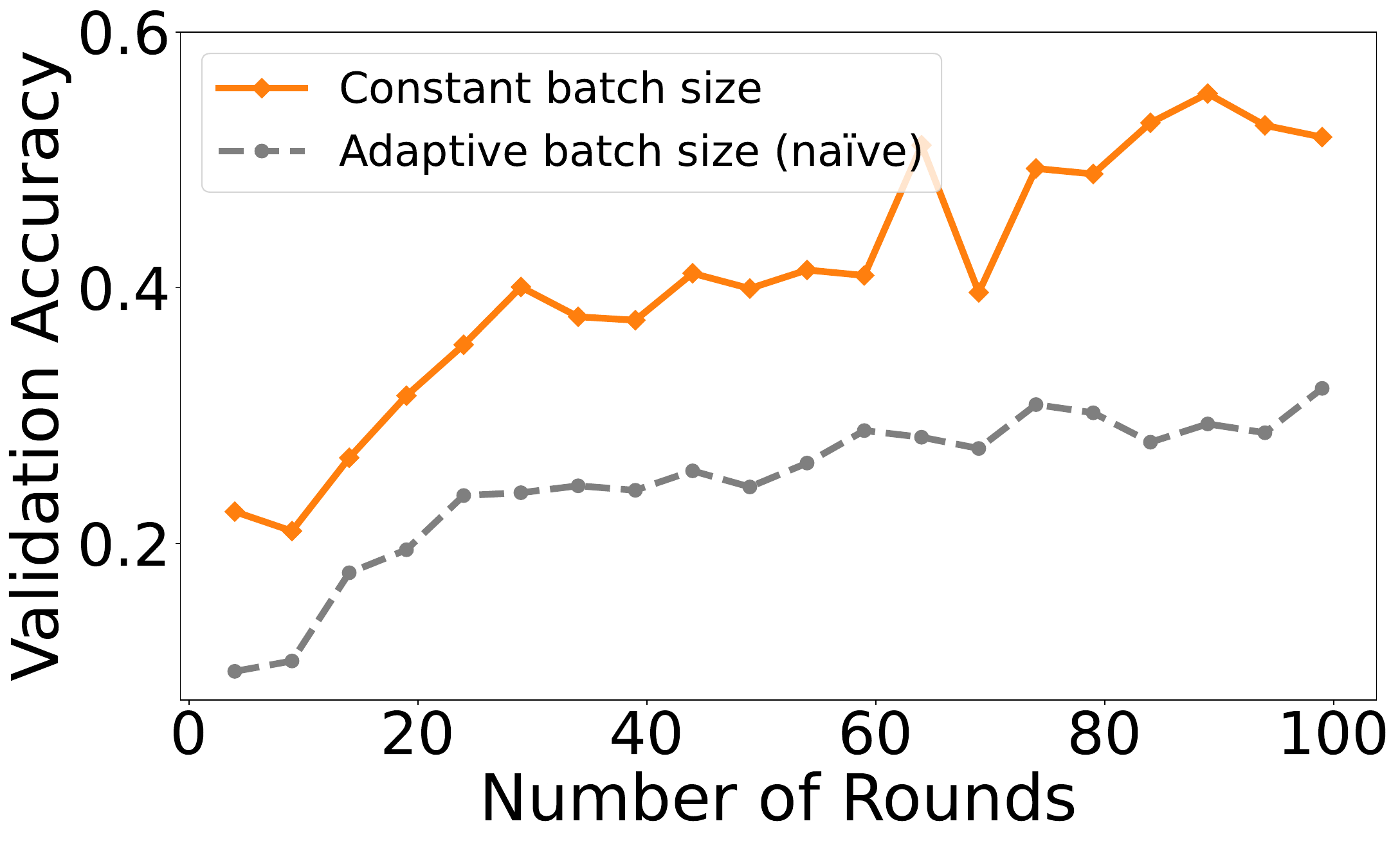}
    \caption{Na\"ively adapting client batch sizes without adapting the number of samples used in each model training round compromises training progress compared to using a constant batch size.}
    \label{fig:motivation_bsz}
\end{figure}

\textbf{Increasing batch sizes can improve system throughput.} 
Traditional FL tasks employ a uniform batch size for all clients, which is typically small to accommodate devices with limited memory. This one-size-fits-all approach assumes a linear relationship between runtime and the number of samples processed, and overlooks the potential of powerful clients, e.g., servers and modern mobile devices, to handle larger batch sizes by training more data samples in parallel, thus achieving higher throughput. We validate this intuition by profiling heterogeneous devices in \S\ref{sec:evaluation}. Customizing batch sizes according to each client's capacity may thus improve the overall system throughput. 
A straightforward method to do so would select the batch size with maximum throughput for each client, without changing the total number of data samples used in the training round. 

\textbf{Increasing the batch size can slow training progress per round.} We evaluate this na\"ive method in training the Cifar10 dataset across 200 clients, with 10 clients randomly chosen to participate in per round. Figure~\ref{fig:motivation_bsz} contrasts the round-to-accuracy performance with a constant batch size of 10 and an adaptive batch size varying between 10 and 100 for different clients. The results indicate a noticeable drop in accuracy at each round round when using na\"ively adaptive batch sizes. As a result, additional rounds are needed to achieve the desired accuracy level,  which potentially increases the overall time-to-accuracy metric.


The reason behind Figure~\ref{fig:motivation_bsz}'s result is that as batch sizes grow, \textit{the learning contribution of each individual training sample diminishes}, leading to less training progress made by clients with larger batches. Consequently, the collective progress across all clients reduces, and the model may even be skewed towards clients making greater statistical progress. To navigate this, it is necessary to increase the number of iterations and thus increase the number of samples used for training for clients with larger batch sizes, which can compensate for their slower statistical progress. However, predicting the exact increment needed to match progress is challenging; moreover, increasing the number of iterations also increases the per-round training time, which may then increase the overall time-to-accuracy. We design methods to navigate these tradeoffs in \S~\ref{sec:system_adapt}.

\begin{figure}[h]
    \centering
    \includegraphics[width=1\linewidth]{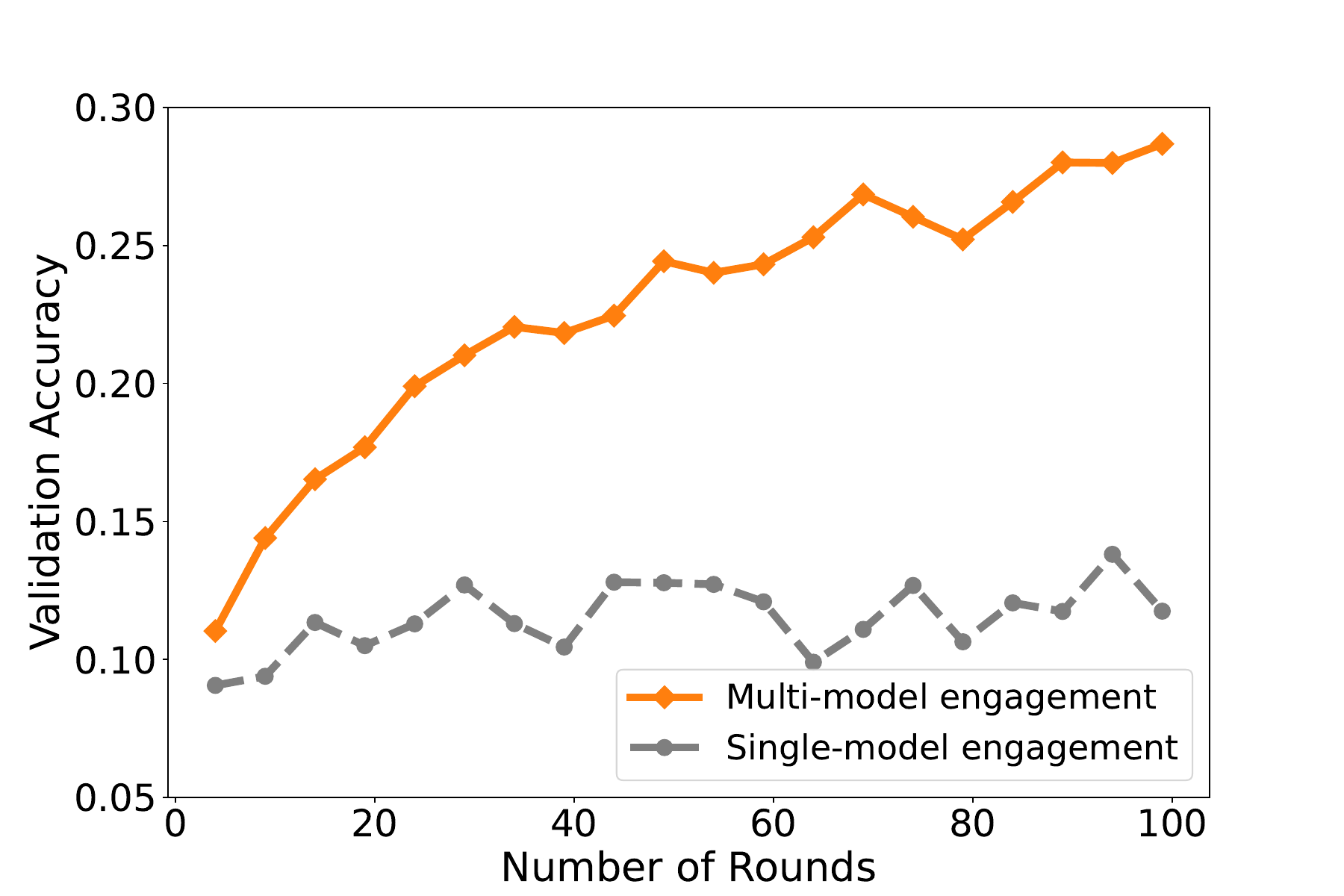}
    \caption{Multi-model engagement, in which some clients may train multiple models in a single global training round, accelerates model training compared to single-model engagement.}
    \label{fig:motivation_multi}
\end{figure}

\subsection{Multi-Model Engagement Per Client}

\label{sec:motivation_multi_task}

Existing studies minimize client idle time by burdening faster clients with larger training size, such as using larger batch sizes or more iterations~\cite{oort}. However, this strategy can inadvertently bias the model towards the data from these faster clients, compromising the final model quality~\cite{cho2020client}. The multi-model scenario presents an alternative to stragglers: \textit{allocating multiple models to faster clients}, thereby engaging more participants per model in each training round. Engaging more clients per training round expands and diversifies the training data, potentially addressing the challenge of non-IID (independent and identically distributed) data across clients.
Specifically, in a non-IID scenario, like one where each client holds data for just one of Cifar10's 10 classes, Figure~\ref{fig:motivation_multi} shows that engaging 2 times more clients per round leads to significantly faster training progress per round than having faster clients train on 2 times more data samples in each round. Consistently allocating the same fast clients to multiple models, however, may bias the model away from slower clients' data~\cite{cho2020client}. We design intelligent client selection methods to control these tradeoffs in $\S\ref{sec:selection}$.
\section{\name{} Overview}
\label{sec:system_overview}

We present \name{}, a multi-model FL training framework, to optimize batch sizes and client selection with multi-model engagement. We next provide an overview of how \name{} fits in the \mmfl{} lifecycle.

In our \mmfl{} scenario, an FL developer sends a job to the parameter server to train a set of models, $\mathcal{\Tilde{M}}$, in parallel, using local data from a set of clients $\mathcal{\Tilde{N}}$. Each client $i\in\mathcal{\Tilde{N}}$ has $M = |\mathcal{\Tilde{M}}|$ local datasets $\mathcal{D}_{i,1}, \mathcal{D}_{i,2},\ldots,\mathcal{D}_{i,M}$ corresponding to the $M$ models, with associated loss function $\ell_{i,j,d}(w_j)$ for each data point $d\in\mathcal{D}_{i,j}$. Here $w_j$ denotes the parameters of model $j\in\mathcal{\Tilde{M}}$. The learning objective is to choose $w_j$ so as to minimize the average loss 
$
\frac{1}{\sum_{i \in \mathcal{\Tilde{N}}} |\mathcal{D}_{i,j}|} \sum_{i \in \mathcal{\Tilde{N}}} \left( \sum_{d \in \mathcal{D}_{i,j}} \ell_{i,j,d}(w_j) \right)
$.


\begin{algorithm}[h!]
  \caption{Pseudo-code of \name{} end-to-end runtime.}\label{pseudo-code}
  \begin{algorithmic}[1]
    \Statex \textbf{Input}: a set of registered clients $\mathcal{\Tilde{N}}$, a set of  models $\mathcal{\Tilde{M}}$ with initial weights $\mathcal{\Tilde{W}} = \{ \Tilde{w_{1}}, \dots, \Tilde{w_{M}} \}$ and target accuracy $\mathcal{\Tilde{A}} = \{ \Tilde{a_{1}}, \dots, \Tilde{a_{M}}\}$, initial batch size $m_{0}$, initial number of iterations $k_{0}$, number of clients to select in each round $s$

    \Statex \textbf{Output}: a list of trained models weights $\mathcal{W} = \{ w_{1}, \dots, w_{M}\}$

    \For{each model $j \in \mathcal{\Tilde{M}}$}
        \State $w_j \xleftarrow{} \Tilde{w_j}$ \Comment{init. model weights}

        \State $a_{j} \xleftarrow{}$ \textbf{Evaluate}($w_j$) \Comment{init. model accuracy}

        \For{each client $i \in \mathcal{\Tilde{N}}$}
            \State $m_{i,j}^{*}=m_{0}$ \Comment{init. batch size}
            \State $k_{i,j}^{*}=k_{0}$ \Comment{init. number of iterations} 
            \State $U_{i,j}=\infty$ \Comment{init. client utilities}
        \EndFor

    \EndFor

    \For{each training round $r = 0,1,2,\ldots$}

        \State $\mathcal{M} \xleftarrow{} \{j | a_{j} < \Tilde{a_{j}},j \in \mathcal{\Tilde{M}}\}$\label{line_active_models} \Comment{models whose target accuracy is not achieved}

        \If{$|\mathcal{M}| = 0$}
            \State \textbf{exit}

        \EndIf

        \State $\mathcal{N} \xleftarrow{} \{ i | \textbf{Available}(i)=1, i \in \mathcal{\Tilde{N}}\}$\Comment{available clients}\label{line_active_clients}

        \State $\mathcal{U} \xleftarrow{} \{U_{i,j} | i \in \mathcal{N}, j \in \mathcal{M}\}$ \Comment{client utilities}

        \State $\mathcal{\Tilde{X}} \xleftarrow{} \{ \Tilde{x}_{i,j} | i \in \mathcal{N}, j \in \mathcal{M}  \}$

        \For{each $\Tilde{x}_{i,j} \in \mathcal{\Tilde{X}} $}\label{line_exclude_start}
            \If{$|\mathcal{D}_{i,j}| > 0$} \Comment{client possesses training data}
                \State $\Tilde{x}_{i,j} = 1$ \Comment{client is considered for selection}
            \Else
                \State $\Tilde{x}_{i,j} = 0$ \Comment{client is not considered for selection}
            \EndIf
        \EndFor\label{line_exclude_end}

        \State $\mathcal{X} \xleftarrow{}$ \textbf{SelectClients}($\mathcal{N}, \mathcal{M}, \mathcal{U}, \mathcal{\Tilde{X}}, s$) \Comment{client selection (\S\ref{sec:selection}})\label{line_select_clients}

        \For{each selected client $i$ s.t. $\sum_{j \in \mathcal{M}} x_{i,j} > 0$ \textbf{parallelly}}\label{line_client_start}

            \For{each assigned model $j$ s.t. $x_{i,j}=1$}\label{line_model_start}

                \State $\mathcal{L}_{i,j}, \nabla \ell_{i,j}, \phi \xleftarrow{}$ \textbf{Train}($w_j, D_{i,j}, m_{i,j}^{*}, k_{i,j}^{*}$)~\Comment{Train the model with optimized batch size and number of iterations}\label{line_train}



                \State $m_{i,j}^{*}, k_{i,j}^{*} \xleftarrow{}$ \textbf{AdaptBatchSize}($m_{0}, k_{0}, \phi$) \Comment{batch size adaptation ($\S\ref{sec:system_adapt}$)}\label{line_adapt}

                \State $U_{i,j}^{*} \xleftarrow{}$ \textbf{UpdateClientUtility($\mathcal{L}_{i,j},m_{i,j}^{*}, k_{i,j}^{*}$)} \Comment{update client utility using the latest sample loss $\mathcal{L}_{i,j}$, batch size $m_{i,j}^{*}$ and number of iterations $k_{i,j}^{*}$ (Equation~\ref{formula_utility})}\label{line_update_utility}

                \State \textbf{Upload}($\nabla \ell_{i,j}, U_{i,j}^{*}$) \Comment{upload the model update and client utility to the server}\label{line_upload}

            \EndFor\label{line_model_stop}
        \EndFor\label{line_client_stop}

        \For{each model $j \in \mathcal{M}$}
            \State $w_{j} \xleftarrow{}$ \textbf{UpdateModel} ($w_{j}$, $\{\nabla \ell_{i,j} | x_{i,j}=1, i \in \mathcal{N} \}$)\label{line_aggregate}

            \State $a_{j} \xleftarrow{}$ \textbf{Evaluate} ($w_{j}$)~\Comment{update model accuracy}\label{line_server_update_acc}

            \For{each selected client $i$ s.t. $x_{i,j}=1$}
                \State $U_{i,j} \xleftarrow{} U_{i,j}^{*}$ \Comment{update client utilities}\label{line_server_update_client_utility}

            \EndFor
            
        \EndFor

    \EndFor
    
  \end{algorithmic}
\end{algorithm}

\textbf{End-to-end runtime.} \name{} is designed to minimize the time-to-accuracy metric for each model, defined as the wall-clock time required to run sufficient training rounds that the target accuracy of each model $\Tilde{a}_{j} \in \mathcal{\Tilde{A}}$ is achieved.
Algorithm~\ref{pseudo-code} outlines how \name{} operates to achieve this objective. 

In each training round, the server first checks which models still need training (line~\ref{line_active_models}) and gathers information about which clients are available to participate (line~\ref{line_active_clients}) in this round.
\modify{Client $i$ is not considered for selection for model $j$ if the client does not have the training data for this model (lines~\ref{line_exclude_start}-~\ref{line_exclude_end}).}
The server selects a subset of clients $\mathcal{X}$ using the multi-model client selection method (\S\ref{sec:selection}), with $x_{i,j} \in \mathcal{X}$ denoting whether or not client $i$ is assigned with model $j$ in this round (line~\ref{line_select_clients}).

Next, all selected clients perform model training over their own datasets in parallel (lines~\ref{line_client_start}-\ref{line_client_stop}), and each single client trains each of its assigned models in sequence (lines~\ref{line_model_start}-\ref{line_model_stop}).
For each assigned model $j$, client $i$ performs training using batch size $m_{i,j}^{*}$ and number of iterations $k_{i,j}^{*}$, which are optimized to minimize training time (\S\ref{sec:system_adapt}), and records the gradient noise scale $\phi$ (to be used for batch size adaptation), model updates $\nabla \ell_{i,j}$ and a list of per-sample loss $\mathcal{L}_{i,j} = \{ l_{i,j,d}(w_{j}) | d \in \mathcal{D}_{i,j} \}$ (to be used for updating client utilities) (line~\ref{line_train}).
After training, the client co-optimizes the batch size $m_{i,j}^{*}$ and number of iterations $k_{i,j}^{*}$ to minimize training time (line~\ref{line_adapt}), using the batch size adaptation method which will be described in \S\ref{sec:system_adapt}. 
The client utility is then updated using the latest sample loss $\mathcal{L}_{i,j}$, batch size $m_{i,j}^{*}$, and number of iterations $k_{i,j}^{*}$ according to Equation~\ref{formula_utility} (line~\ref{line_update_utility}).
Finally, in parallel to training the next assigned model, the clients uploads the models updates and client utility to the central server (line~\ref{line_upload}).

After all the clients have finished training all the assigned models, the server then aggregates the updates $\nabla \ell_{i,j}$ for each allocated (client, model) pair as in the FedAvg algorithm~\cite{fedavg} (line~\ref{line_aggregate}). Other aggregation methods can be used without changing our learning framework.
At the end of each round, the server updates the model accuracy (line~\ref{line_server_update_acc}) and client utilities (line~\ref{line_server_update_client_utility}) before advancing to the next round.

\begin{figure}[h]
    \centering
    \includegraphics[width=1\linewidth]{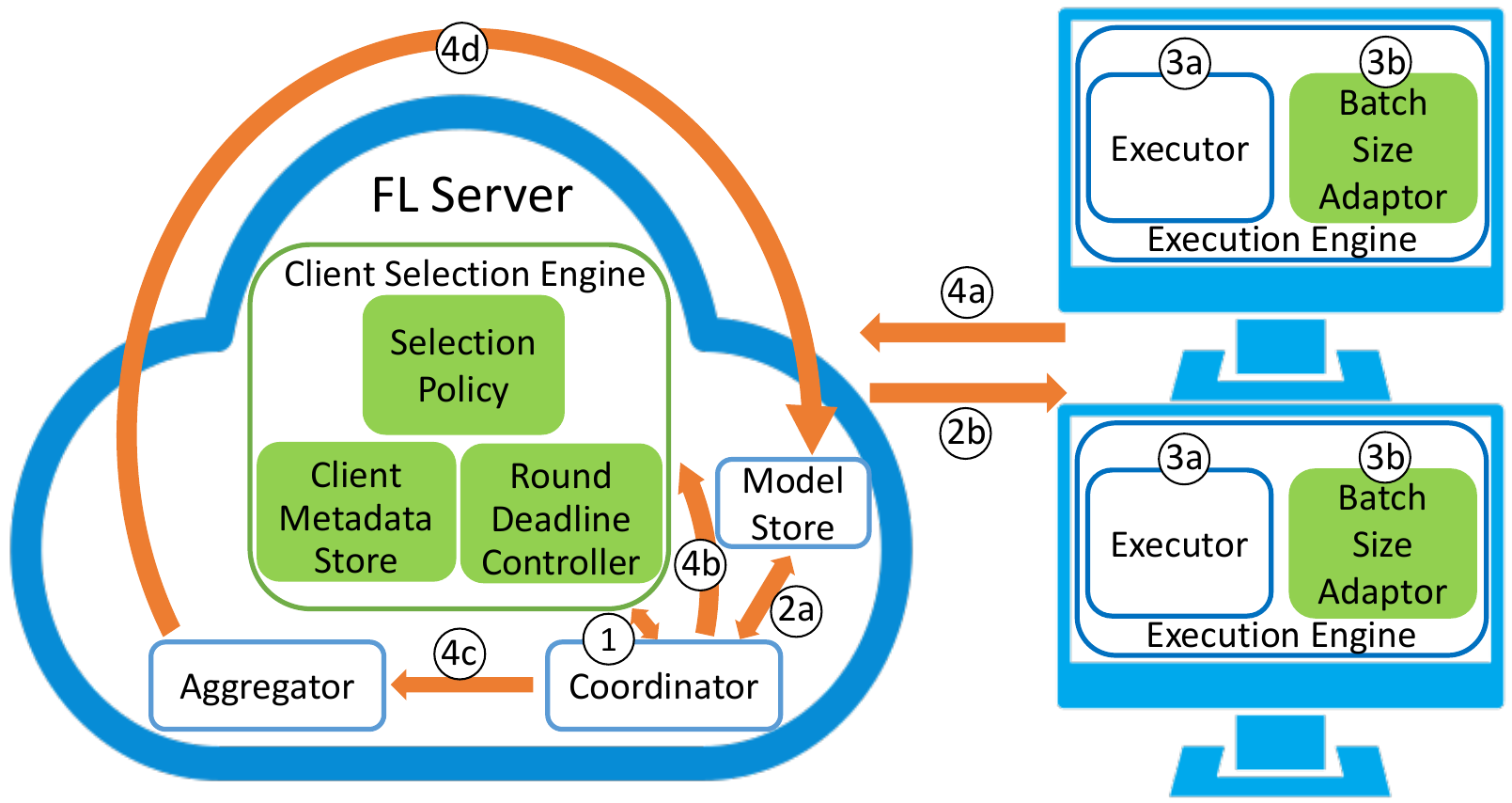}
    \caption{\name{} architecture and operation flow in each training round.}
    \label{fig:diagram}
\end{figure}

\textbf{System components.}
\name{} consists of two primary components: a batch size adaptor that adapts the batch size for each client (Section~\ref{sec:system_adapt}) and a client selection engine that allocates client to models (Section~\ref{sec:selection}). To get started, FL developers provide model specifications along with key parameters such as the client count $S$, batch size $m_0$, and number of local iterations $k_0$. 

Figure~\ref{fig:diagram} illustrates \name{}'s operation flow in each training round:

\begin{enumerate}
    \item \textit{Client selection:} the coordinator queries the client selection engine to allocate clients to each model (line~\ref{line_select_clients}).
    \item \textit{Task distribution:} (a)  the coordinator retrieves the current global models and (b) distributes them to their chosen clients. 
    \item \textit{Execution:} selected clients (a) train their allocated models and update the gradient noise scale (line~\ref{line_train}), and (b) update gradient noise scale and adapt their batch sizes for each model, to be used in the next training round (line~\ref{line_adapt}).
    \item \textit{Result aggregation:} each client (a) transmits its model updates and revised client utility, post-batch size adjustments (line~\ref{line_upload}). Subsequently, the coordinator (b) updates client utilities (line~\ref{line_server_update_client_utility}), (c) aggregates model updates (line~\ref{line_aggregate}) and evalutes the model accuracy (line~\ref{line_server_update_acc}), and (d) saves the updated global model.
\end{enumerate}


Notice that the client selection engine according to the utility which is reported in the client's last participated round.
Since the client utility depends on the batch size, adapting the batch size at a round's start could misalign with this reported utility. Hence, we adapt the batch size post-training.

\section{\name{} Design}
\label{sec:system_design}

In this section, we explain how \name{} optimizes batch sizes (\S\ref{sec:system_adapt}) and selects clients (\S\ref{sec:selection}). We also describe the implementation of \name{} (\S\ref{sec:impl}).

\subsection{Batch Size Adaptation}
\label{sec:system_adapt}
\name{} aims to adjust each client's batch size to optimize system throughput. As highlighted in Section~\ref{sec:motivation_adapt}, increasing the batch size can compromise convergence due to diminished statistical progress post-adjustment. For effective throughput improvement without compromising statistical quality, three challenges arise: (i) quantifying statistical progress, (ii) preserving this progress post-batch size adjustment, and (iii) selecting the optimal batch size for each client. We discuss how we address each of the challenges below. The details are shown in Algorithm~\ref{alg}.

\begin{algorithm}[h]
  \caption{Function AdaptBatchSize.}\label{alg1}
  \begin{algorithmic}[1]
    \Statex Input: $m_0, k_0, \phi$
    \Statex Output: $m*, k*$
    \State $ m^{*} = \argmax_{m} \theta(m) \cdot \frac{\phi+m_0}{\phi+m} \cdot \varphi(m_0)$ \Comment{Adapt the batch size.}
    \State $ k^{*}  = \left\lceil \frac{m_0}{m^{*}} \cdot \frac{\phi+m_0}{\phi+m^{*}} \cdot k_0  \right\rceil$\label{alg:iter} \Comment{Adapt the number of iterations after adapting the batch size.}
  \end{algorithmic}
  \label{alg}
\end{algorithm}

\textbf{How to quantify statistical progress?}
We define the \textit{statistical progress} $\sigma(m, k)$ as the amount of progress a client can make by training the model on its local data samples using batch size $m$ for $k$ local iterations. Statistical efficiency $\varphi(m)$ represents the amount of progress that can be made with each training sample using batch size $m$. Thus, the statistical progress is: $\sigma(m, k) = k \cdot m \cdot \varphi(m)$.

To quantify $\varphi(m)$, we leverage previous work~\cite{johnson2019adascale, mccandlish2018empirical} relating the statistical efficiency to the Gradient Noise Scale (GNS) $\phi$, which measures the noise-to-signal ratio of the stochastic gradient. A larger GNS implies a smaller decrease in the statistical efficiency from increasing training parameters such as the batch size and learning rate. The GNS can vary greatly between different DL models~\cite{golmant2018computational}. It is also non-constant and tends to gradually increase during training, by up to 10× or more~\cite{mccandlish2018empirical}. Thus, it is possible to attain significantly better statistical efficiency for large batch sizes in later training rounds.

According to a previous study on large-batch training~\cite{mccandlish2018empirical}, the statistical efficiency using batch size $m$, relative to that when using the initial batch size $m_0$, can be captured by:
\begin{equation}
    \frac{\varphi(m)}{\varphi(m_0)} = \frac{\phi+m_0}{\phi+m}
\end{equation}
Thus, the statistical progress that can be made when using batch size $m$ and $k$ iterations, relative to that when using $m_0$ and $k_0$, can be expressed as:
\begin{equation}
    \frac{\sigma(m, k)}{\sigma(m_0, k_0)} = \frac{m \cdot k \cdot \varphi(m)}{m_0 \cdot k_0 \cdot \varphi(m_0)} = \frac{m \cdot k}{m_0 \cdot k_0} \cdot \frac{\phi+m_0}{\phi+m}
\end{equation}
\textbf{How to maintain statistical progress after adapting batch sizes?}
Increasing a client's batch size from $m_0$ to $m$ and proportionally reducing the number of iterations from $k_0$ to $k$ to use the same amount of training data, i.e., $m \cdot k = m_0 \cdot k_0$, results in diminished statistical progress since $\frac{\phi+m_0}{\phi+m} \le 1$.
This approach then requires more training rounds to achieve equivalent model accuracy, incurring additional training time and communication overhead. Furthermore, the model becomes skewed towards clients making more substantial progress each round~\cite{cho2020client}, resulting in inferior model quality. This effect contributes to the poor time-to-accuracy performance observed with adaptive batch sizes in Figure~\ref{fig:motivation_bsz}.
To uphold the same level of statistical progress when a client adapts its batch size, we adapt the number of iterations $k$ according to line 3 in Algorithm~\ref{alg}.

\textbf{How to optimize batch size?}
\label{sec:policy}
Since we increase the number of local iterations for larger batches in order to maintain statistical progress, selecting the batch size for a client based solely on maximizing throughput, as suggested in Section~\ref{sec:motivation_adapt}, may not actually minimize the client's round execution time. 
Thus, similar to the strategy proposed in~\cite{pollux}, we instead choose the batch size $m^{*}$ that maximizes the statistical progress per second, thereby minimizing execution time (line 2 in Algorithm~\ref{alg}).
Here, the throughput $\theta(m)$ for a client using batch size $m$ denotes the number of samples processed per second (samples/sec), while the statistical efficiency $\varphi(m)$ indicates the progress achievable per sample (progress/sample). Thus, $\theta(m) \cdot \varphi(m)$ quantifies the progress achievable per second (progress/sec). 
We formulate the batch size adaptation problem as \textbf{P1}. \modify{Since \textbf{P1} has a single, discrete optimization variable (the batch size $m$), we simply solve \textbf{P1} by iterating over a range of possible batch sizes (10-100 used in \S\ref{sec:evaluation}).}

\begin{align}
    \textbf{P1}: ~~~~~& \max_{m} \theta(m) \cdot \varphi(m)\label{p1_obj}\\
    ~~~~~& s.t. \varphi(m) = \frac{\phi+m_0}{\phi+m} \cdot \varphi(m_0)
\end{align}

\subsection{Multi-Model Client Selection}\label{sec:selection}
As detailed in Section~\ref{sec:motivation_multi_task}, allocating multiple models to individual clients, which we call \textit{multi-model engagement}, enabled by our MMFL scenario, proves more efficient in reducing idle time than increasing the data trained at faster clients as proposed in single-model FL work. 
We elaborate our multi-model client selection technique below, which optimizes the clients selected for each model and encourages multi-model allocation per-client.

\textbf{Tradeoff between system throughput and data quality.}
Our goal is to select clients so as to optimize our FL models' time-to-accuracy, which depends on two aspects of the selected clients and current model: (i) \textit{system throughput:} the duration of each training round, given this client's chosen batch size for each model; and (ii) \textit{data quality:} the degree to which the current model aligns with the client's local data. We quantify both to define the \textit{utility} of selecting each client for each model in a given training round.

Previous work on client selection~\cite{eds_multi_aaai, eds_multi_journal} simply models a client's data quality as the frequency with which this client has been selected. However, this approach cannot effectively capture clients' data quality~\cite{oort}. An ideal way to model a client's data quality should efficiently capture the reduction in model loss from including the client's data in the training.
Following the method from~\cite{oort}, we thus define the data quality $U^{data}_{i,j}$ of client $i$ for model $j$ as:
\begin{equation}
    U^{data}_{i,j} = |B_{i,j}| \cdot \sqrt{\frac{1}{|B_{i,j}|} \cdot \sum_{b \in B_{i,j}} \mathcal{L}(b)^2}, \label{formula_data_utility}
\end{equation}
where $B_{i,j}$ is the subset of training data used by client $i$ to train the model $j$ and $\mathcal{L}(b)$ is the sample loss. Furthermore, the system throughput $U_{i,j}^{sys}$ of client $i$ for model $j$ is characterized by:

\begin{equation}
    U_{i,j}^{sys} = \frac{D}{t_{i,j}} \label{formula_system_utility}
\end{equation}

where, $D$ is the deadline of this round (we will discuss how we select the deadline over time later) and $t_{i,j}$ is the execution time given client $i$'s adapted batch size using Algorithm~\ref{alg}. Intuitively, the faster a client can finish a job before the deadline, the higher its system utility. Combining system throughput and data quality, we define the \textit{utility} $U_{i,j}$ of client $i$ for model $j$ as:

\begin{equation}
    U_{i,j} = \overline{U}_{i,j}^{sys} \cdot \overline{U}^{data}_{i,j}\label{formula_utility}
\end{equation}

,where $\overline{U}_{i,j}^{sys}$ and $\overline{U}^{data}_{i,j}$ represent the system throughput and data quality, normalized across all clients for model $j$ respectively.

\textbf{Staleness in estimating client utility.}
Note that while our utility function $U_{i,j}$ is designed to capture a client's contribution to the training process, we do not have up-to-date utility values for all clients during training, primarily due to two factors. 

First, the data qualities become stale for clients not selected in recent rounds, as their loss calculations are based on an older model, not accounting for recent updates. 
Second, the gradient noise scale, $\phi$, typically increases~\cite{mccandlish2018empirical} during training. As this happens, increasing the batch size does not significantly diminish statistical efficiency as $\lim_{\phi \rightarrow \infty} \frac{\phi+m_0}{\phi+m} = 1$. Since we use the batch size determined in the client's \textit{previous} training round for this model to estimate the system throughput, which is based on a smaller value of $\phi$, our system throughput estimate may be outdated; clients might achieve lower execution times with larger batch sizes without compromising the model's statistical integrity.

Similar to the strategy in work~\cite{oort}, we add an uncertainty term $\alpha \cdot \sqrt{\frac{R}{r_{i,j}}}$ to $U_{i,j}$, which shares the same shape of the confidence in bandit solutions~\cite{lin2022neural}, to account for these staleness factors, where $R$ is the current round and $r_{i,j}$ is the number of rounds client $i$ has been selected for model $j$. Intuitively, we gradually increase the utility of a client if it has not been selected to train this model for a long time.

\textbf{Client selection objective.}
We define the utility $U_{j}$ of a model $j$ as the combined utilities of the clients allocated to it. Our goal is to optimize the combined utilities across all models, thereby prioritizing high-utility clients as well as encouraging multi-model allocation per client (Equation~\ref{obj}). We formulate the client selection problem as $\textbf{P2}$.

\begin{align}
     \textbf{P2}: \max_{x_{i,j}} ~~~~~& \sum_{j \in \mathcal{M}} \sum_{i \in \mathcal{N}} x_{i,j} \cdot \left( U_{i,j} + \sqrt{\frac{R}{r_{i,j}}} \right)\ \label{obj} \\
    s.t. ~~~~~& \sum_{j \in \mathcal{M}} x_{i,j} \cdot t_{i,j} \le D, ~~~~~~\forall i \in \mathcal{N} \label{cons1} \\
        & \sum_{i \in \mathcal{N}} \mathds{1}\left(\sum_{j \in \mathcal{M}} x_{i,j}\ge 1 \right) = S \label{cons2}\\
        & x_{i,j} \le \Tilde{x}_{i,j}, ~~~~~~\forall i \in \mathcal{N} j \in \mathcal{M} \label{cons3}
\end{align}

$x_{i,j}\in\{0,1\}$ indicates the allocation of client $i$ to model $j$. Constraint~\ref{cons1} ensures that all clients complete their allocated tasks within a set deadline (the adaptation of this deadline is discussed later). Constraint~\ref{cons2} confirms that the requisite number of clients are chosen, with $S$ representing the necessary client count and $\mathds{1}\left(\sum_{j \in \mathcal{M}} x_{i,j} \ge 1 \right)$ serving as an indicator function that signifies whether a client is allocated with at least one model.
\modify{Constraint~\ref{cons3} ensures that client $i$ will not be selected for model $j$ if the client does not have the corresponding training data. Here, $\Tilde{x}_{i,j}$ denotes whether client $i$ possesses training data for model $j$.}

\textbf{Dynamic deadline control.} Equation~\ref{obj} incentivizes maximizing the number of models per client to enhance collective model utilities. While engaging more clients per model can accelerate round-to-accuracy with richer, more diverse data, the marginal gains diminish over time~\cite{logfair}. Moreover, allocating more models to each client prolongs round duration, potentially increasing the time-to-accuracy. To balance these factors, we adjust the round deadline dynamically based on the moving average of validation loss: increasing it if the average rises, and decreasing it if it falls, similar to the approach in work~\cite{fedbalancer}.

Assuming $T = \{t_{i,j} | i \in \mathcal{N}, j \in \mathcal{M}\}$ denotes the execution times for any given client-model pair, the deadline is dictated by percentile $p$: $D$ is set at the $p^{th}$ percentile of $T$. Initially, with $p=100$, the deadline is the maximum value in $T$. \name{} assesses the appropriateness of the current deadline, $G$, using $G_D = \frac{L_{test}}{D}$, where $L_{test}$ signifies average test loss. This $G_{D}$ is then accumulated in a list $G$. \name{} contrasts the sum of $G_D$ values from previous $w$ rounds, $\sum G[R-2w : R-w]$, against those from more recent $w$ rounds, $\sum G[R-w : R]$. If earlier rounds surpass recent ones, indicating stable training, \name{} decreases $p$ by $\epsilon$, aiming to shorten round duration and decreases $p$ by $\epsilon$ otherwise. Conversely, if recent rounds are higher, suggesting the need for involving more clients, \name{} increases $p$ by $\epsilon$.

\textbf{Overlapping communication with computation.} Allocating several models to each client per round will not incur additional communication overhead due to smart overlapping: it can train the current model while concurrently transmitting updates for the previous model in the background.


\textbf{Complexity analysis.}
\textbf{P2} can actually be considered as a Multi Knapsack Problem, and is thus NP-hard~\cite{pisinger1999exact}.
We re-formulate constraint~\ref{cons2}  to transform \textbf{P2} into an ILP (Integer Linear Program) by defining appropriate auxiliary variables $\mathds{1}_{i}, l_{i}$ as follows.

\begin{align}
    & l_{i} = \sum_{j \in \mathcal{M}} x_{i,j}, \forall i \in \mathcal{N}\\
    & \mathds{1}_{i} \le l_{i}, \forall i \in \mathcal{N}\\
    & \mathds{1}_{i} \cdot M \ge l_{i}, \forall i \in \mathcal{N}
\end{align}

We then implement a solver for \textbf{P2} in Gurobi~\cite{gurobi}, leveraging its advanced algorithms and parallel processing capabilities for faster solving times. Exact solutions to \textbf{P2} can be efficiently found less than 2 seconds for our (fairly realistic) scenarios (selecting 60 out of 200 clients).

\subsection{Implementation}
\label{sec:impl}

Existing FL benchmark platforms~\cite{lai2022fedscale, he2020fedml} only support the single-job setting. We built \textit{the first benchmark platform for MMFL} and implemented \name{} in our platform with 2984 lines of Python code. (1) The platform has an \textit{automated batch size adaptation mechanism} so that users can try different approaches on top of adaptive batch sizes. We achieved this by integrating AdaptDL~\cite{pollux} library to log the gradient noise scale, superseded its batch size optimization with ours, and added functionality of co-adapting batch sizes and the number of iterations.
(2) Our platform \textit{supports various data partition strategies} such as IID, shard~\cite{fedavg}, and Dirichlet~\cite{yurochkin2019bayesian}, by extending FedLab~\cite{zeng2021fedlab}.
(3) We designed a \textit{cluster backend}, which automates and parallelize training of across the available cluster CPUs and GPUs with the support of GPU/CPU sharing.
(4) Our system is designed for versatility, supporting \textit{user-provided system throughput traces and data-client mappings}, allowing for evaluation in diverse scenarios.
(5) The framework offers \textit{user-friendly APIs} which enable the seamless transition of single-model client selection strategies to multi-job settings, as well as the development of customized multi-job client selection strategies.
(6) Our system is \textit{robust}; in the event of disruptions, it auto-saves and loads the most recent checkpoint to ensure continuity.



\section{Evaluation}\label{sec:evaluation}

\modify{In this section, we describe the experimental setup in \S\ref{eval:setup}. Next, we compare the time-to-accuracy performance and final model quality in \S\ref{sec:end_to_end_perf}. We further analyze how \name{} optimizes batch sizes for different models and device types, study the impact of batch size adaptation and model-client selection on performance improvement through ablation studies, and evaluate how \name{} ensures model fairness in \S\ref{eval:closer}. Finally, we investigate the performance of \name{} with different parameters in \S\ref{eval:sensitivity}.}

\subsection{Experiment Setup}\label{eval:setup}
\textbf{Datasets and models.}
We evaluate our framework using nine real-world datasets of varying scales with well-known models, grouped into sets of three for parallel execution. Table~\ref{tab:datasets} shows the datasets and models in each group. We refer to SquadV1 with Bert as SquadV1-1, and SquadV1 with DistilBert as SquadV1-2.



\begin{table*}[h]
    \centering
    \begin{tabular}{|c|c|c|c|c|c|} 
    \hline 
    Group & Dataset & Model & Number of Samples & Model Size (parameters) & Task \\ 
    \hline 
    \multirow{3}{*}{A} 
    & Fashion Mnist & CNN & 70,000 & 1.5M & Image Classification \\ 
    \cline{2-6} 
    & Cifar-10 & ResNet18 & 60,000 & 11.2M & Image Classification \\ 
    \cline{2-6} 
    & Google Speech & MobileNetV2 & 104,667 & 3.4M & Speech Recognition \\ 
    \hline 
    \multirow{3}{*}{B} 
    & Celeba & ResNet18 & 30,000 & 11.2M & Image Classification \\ 
    \cline{2-6} 
    & Mnist & CNN & 70,000 & 1.2M & Image Classification \\ 
    \cline{2-6} 
    & Cifar-100 & ResNet34 & 60,000 & 21.8 & Image Classification \\ 
    \hline 
    \multirow{3}{*}{C} 
    & SquadV1 & BERT & 100,000 & 110M & Question Answering \\ 
    \cline{2-6} 
    & SquadV1 & DistilBERT & 100,000 & 66M & Question Answering \\ 
    \cline{2-6} 
    & SquadV2 & BERT & 150,000 & 110M & Question Answering \\ 
    \hline 
    \end{tabular}
    \caption{Datasets, models, number of samples, model sizes, and tasks.}
    \label{tab:datasets}
\end{table*}

\textbf{Heterogeneous system throughput and data quality.}
We employ 32 NVIDIA Tesla T4 GPUs to simulate a FL setting with a central server and 200 clients. The heterogeneous system throughput is emulated using AI Benchmark data~\cite{ignatov2019ai}, which documents device runtimes for various models with a single batch size. To extrapolate runtime variations for batch sizes between 10 and 100, we profile three device types: a GPU (NVIDIA Tesla T4), CPU (Intel Xeon Platinum 8259Cl), and mobile device (Raspberry Pi Model 4). Each client is randomly assigned one of these device types, and its runtime scaling is simulated based on the device-specific profile. Clients receive portions of the real datasets via a Dirichlet distribution~\cite{li2022federated}, resulting in variations in data volume, output distribution, and input features.


\textbf{Baselines.}
We compare with the following six baselines.

\begin{enumerate}
    \item FedAvg~\cite{fedavg}: Engages clients at random. 
    \item FedBalancer~\cite{fedbalancer}: Chooses clients randomly, while shortening training round durations by curtailing the training data for the chosen clients.
    \item Oort~\cite{oort}: Selects clients considering both system throughput and data quality.
    \item RoundRobin~\cite{round_robin}: Randomly sorts clients into $M$ groups per round, assigning each group to one of the $M$ models in a uniformly randomized manner.
    \item Logfair~\cite{logfair}: Maximizes the sum of the logarithmic values of the client counts allocated to each model.
    \item Eds~\cite{eds_multi_aaai}: Allocates clients based on an evaluation of their cross-model system throughput and data quality.
\end{enumerate}


Baselines 1-3 are single-model client selection strategies. We extend these strategies to the multi-model context by repeating the client selection algorithm for each of the $M$ models. Baselines 4-6 are state-of-the-art MMFL works.

\textbf{Methods.} 
We ran FedAvg on each dataset until convergence and subsequently ran experiments with \name{} and other baselines for an equivalent number of rounds. We set the target accuracy to be the lowest final model quality achieved by all the methods. We measured the \textit{clock time that each method  needs to reach the target accuracy} based on client runtime traces, as well as the \textit{final model accuracy}.

\textbf{Configurations.}
As in prior work~\cite{fedbalancer, oort}, we set the initial batch size to 10 and the number of iterations to 20 for all datasets. We use a learning rate of 0.01 for Google Speech, 0.00003 for SquadV1 and SquadV2, and 0.001 for the other datasets. In each round, we select 10 clients for every dataset.
 
\begin{table*}[h]
    \centering
\begin{tabular}{|c|c|c|c|c|c|c|}
\hline
Group & \multicolumn{6}{c|}{A} \\
\hline
Dataset (Round) & \multicolumn{2}{c|}{Fashion Mnist (200)} & \multicolumn{2}{c|}{Cifar-10 (400)} & \multicolumn{2}{c|}{Google Speech (400)} \\
\hline
Method                             & Time         & Acc.       & Time        & Acc.         & Time         & Acc.         \\ \hline
FedAvg                             & 10.04 ± 0.28 & 88.23 ± 0.12 & 20.60 ± 1.31 & 76.22 ± 0.08 & 18.82 ± 1.91 & 60.13 ± 0.35 \\
Oort                               & ~~7.19 ± 0.54  & 88.33 ± 0.25 & 17.21 ± 0.83 & 77.90 ± 0.89 & 16.01 ± 2.22 & 59.80 ± 0.72 \\
LogFair                            & 11.18 ± 0.32 & 88.40 ± 0.36 & 23.42 ± 0.63 & 78.33 ± 0.15 & 18.64 ± 1.50 & 62.37 ± 1.25 \\
Eds                                & ~~6.12 ± 0.79  & 88.00 ± 0.26 & 13.33 ± 0.42 & 76.90 ± 0.36 & 13.04 ± 2.33 & 59.87 ± 1.69 \\
Fedbalancer                        & 11.34 ± 1.70 & 88.70 ± 0.26 & 26.65 ± 3.85 & 77.47 ± 1.72 & 19.89 ± 0.71 & 61.33 ± 4.73 \\
Round Robin                        & 10.05 ± 0.25 & 88.13 ± 0.31 & 20.55 ± 0.55 & 76.67 ± 0.58 & 18.96 ± 2.36 & 60.70 ± 0.78 \\ \hline
\name{}                         & ~~\textbf{2.85 ± 0.27}  & \textbf{89.30 ± 0.05} & ~~\textbf{5.14 ± 0.25} & \textbf{80.10 ± 1.08} & ~~\textbf{5.76 ± 0.82} & \textbf{64.37 ± 0.81} \\ \hline
\end{tabular}


\begin{tabular}{|c|c|c|c|c|c|c|}
\hline
Group & \multicolumn{6}{c|}{B} \\
\hline
Dataset (Round) & \multicolumn{2}{c|}{Celeba (80)} & \multicolumn{2}{c|}{Mnist (100)} & \multicolumn{2}{c|}{Cifar-100 (500)} \\
\hline
Method       & Time        & Acc.        & Time        & Acc.        & Time         & Acc.      \\ \hline
FedAvg       & ~~1.06 ± 0.18 & 94.83 ± 0.38 & ~~3.37 ± 0.27 & 94.43 ± 0.32 & 17.83 ± 0.43 & 50.77 ± 0.32 \\
Oort         & ~~0.99 ± 0.15 & 94.17 ± 0.12 & ~~2.96 ± 0.16 & 94.33 ± 0.06 & 17.40 ± 0.29 & 50.57 ± 0.12 \\
LogFair      & ~~1.46 ± 0.12 & 94.43 ± 0.21 & ~~3.83 ± 0.19 & 94.67 ± 0.06 & 19.00 ± 0.39 & 50.30 ± 0.44 \\
Eds          & ~~0.75 ± 0.16 & 94.67 ± 0.15 & ~~2.32 ± 0.19 & 94.60 ± 0.10 & ~~9.99 ± 0.35 & 50.50 ± 0.35 \\
Fedbalancer  & ~~2.36 ± 0.06 & 85.63 ± 1.12 & ~~2.18 ± 0.37 & \textbf{95.40 ± 0.17} & 11.56 ± 0.93 & 53.10 ± 0.52 \\
Round Robin  & ~~1.19 ± 0.09 & 94.53 ± 0.06 & ~~3.49 ± 0.08 & 94.37 ± 0.23 & 17.83 ± 1.01 & 50.77 ± 0.76 \\ \hline
\name{}      & ~~\textbf{0.68 ± 0.04} & \textbf{95.03 ± 0.06} & ~~\textbf{1.33 ± 0.33} & 95.20 ± 0.10 & ~~\textbf{6.21 ± 0.41}  & \textbf{55.67 ± 0.06} \\ \hline
\end{tabular}

\begin{tabular}{|c|c|c|c|c|c|c|}
\hline
Group & \multicolumn{6}{c|}{C} \\
\hline
Dataset (Round) & \multicolumn{2}{c|}{SquadV1-1 (100)} & \multicolumn{2}{c|}{SquadV1-2 (100)} & \multicolumn{2}{c|}{SquadV2 (200)} \\
\hline
Method       & Time        & Acc.        & Time        & Acc.        & Time         & Acc.      \\ \hline
FedAvg       & ~~6.87 ± 0.20 & 84.40 ± 0.20 & ~~7.54 ± 0.69 & 76.10 ± 0.12 & 17.51 ± 0.79 & 69.40 ± 0.26 \\
Oort         & ~~6.98 ± 0.08 & 83.63 ± 0.25 & ~~6.98 ± 0.08 & 76.17 ± 0.25 & 12.74 ± 1.58 & 70.37 ± 0.32 \\
LogFair      & ~~8.11 ± 0.72 & 84.43 ± 0.06 & ~~8.11 ± 0.72 & 76.73 ± 0.17 & 16.12 ± 0.45 & 70.20 ± 0.36 \\
Eds          & ~~4.42 ± 0.69 & 84.60 ± 0.36 & ~~5.11 ± 0.65 & 76.77 ± 0.25 & ~~9.27 ± 0.75 & 69.87 ± 0.90 \\
Fedbalancer  & ~~6.73 ± 0.70 & \textbf{86.33 ± 0.25} & ~~6.21 ± 0.22 & \textbf{79.87 ± 0.17} & 11.44 ± 5.01 & \textbf{73.97 ± 0.31} \\
Round Robin  & ~~6.53 ± 0.48 & 84.60 ± 0.10 & ~~7.19 ± 1.05 & 76.40 ± 0.42 & 16.38 ± 1.45 & 69.93 ± 0.50 \\ \hline
\name{}      & ~~\textbf{0.81 ± 0.03} & 85.83 ± 0.32 & ~~\textbf{0.98 ± 0.04} & 79.27 ± 0.06 & ~~\textbf{2.25 ± 0.40}  & 73.90 ± 0.10 \\ \hline
\end{tabular}

\caption{\name{} achieves the best time-to-accuracy (hours) performance on all 9 datasets in Table~\ref{tab:datasets}, and the highest final model accuracy (\%) on 5 datasets and second highest final model accuracy on the remaining 4 datasets.}
    \label{tab:1}
\end{table*}

\subsection{End-to-End Performance}
\label{sec:end_to_end_perf}

\textbf{Time-to-accuracy performance.}
Table 1 shows the runtime of \name{} on six datasets compared with the baseline methods. We observed that \name{} shows improved
time-to-accuracy performance over the baselines on every dataset: \name{} achieves  2.2-4.0$\times$, 2.6-5.2$\times$, and 2.3-3.5$\times$ speedup on the Fashion MNIST, Cifar-10, and Google Speech dataset in group A, 1.1-3.5$\times$, 1.6-2.8$\times$, and 1.6-3.1$\times$ speedup on the Celeba, MNIST, and Cifar-100 dataset in Group B, and 5.5-10.0$\times$, 5.2-8.3$\times$, and 4.1-7.8$\times$ speedup on the SquadV1-1, SquadV1-2, and SquadV2 dataset in Group C.


FedAvg achieves poor time-to-accuracy performance because it is agnostic to system throughput and data quality.
FedBalancer shows a higher time-to-accuracy performance on Group B and C compared with FedAvg, but shows an extremely low performance on Group A. The reason behind this is that FedBalancer adopts the epoch training framework, where each client need to complete at least one full epoch.
Note that Google Speech contains 104,667 training samples, and although FedBalancer can reduce the epoch training size by only training over the samples with above-threshold loss values, there are still a larger number of below-threshold ones, resulting in long round time. This is the driving reason why we resort to batch size adaptation to optimize system throughput and reduce round duration, as discussed in \S~\ref{sec:motivation_adapt}.
Even if Oort considers client system throughput and data quality, it is only 1.1-1.2$\times$ faster than FedAvg because it cannot capture clients' cross-model utility, leading to sub-optimal selection solution.
By adopting a cohesive control across different models, EDS, the state-of-the-art MMFL client selection framework, achieves the best performance among the baselines. However, it is still 1.1-5.5$\times$ slower than \name{}, which demonstrates the power of batch size adaptation and multi-model engagement as EDS only considers constant batch size and single-model assignment for each client.


\textbf{Final accuracy.} We observed that \name{} improves time-to-accuracy performance without sacrificing final model accuracy: \name{} achieves up to 1.3\%, 3.9\%, 4.6\%, 0.8\%, and 5.4\% higher final accuracy on the Fashion MNIST, Cifar-10, Google Speech, Celeba, and Cifar-100 dataset respectively. While FedBalancer does achieve 0.2\%, 0.5\%, 0.6\%, 0.07\% higher accuracy on MNIST, SquadV1-1, SquadV1-2, and SquadV2 respectively, its performance is unstable—its final accuracy is 9.4\% lower on the Celeba dataset compared to \name{}.
Furthermore, we found that \name{} has not yet reached convergence. We believe that with additional rounds, \name{} will at least match the final accuracy of FedBalancer while using less time.

\modify{
The final accuracies we achieve align with values reported in prior studies~\cite{zhao2018federated, reddi2020adaptive, devlin2018bert}, though one can change the model and data distribution to improve the baseline final accuracy. Our methods for batch size adaptation and client selection are agnostic to such changes.
In our experiments, we intentionally chose a challenging setup with low client participation and a highly non-IID data distribution, which effectively tests whether our batch size adaptation and client selection methods ensure uniform training progress among clients with largely diverse data.
}

\subsection{A Closer Look at \name{}'s Solution}\label{eval:closer}

\begin{figure*}[h]
  \centering
  \includegraphics[width={\linewidth}]{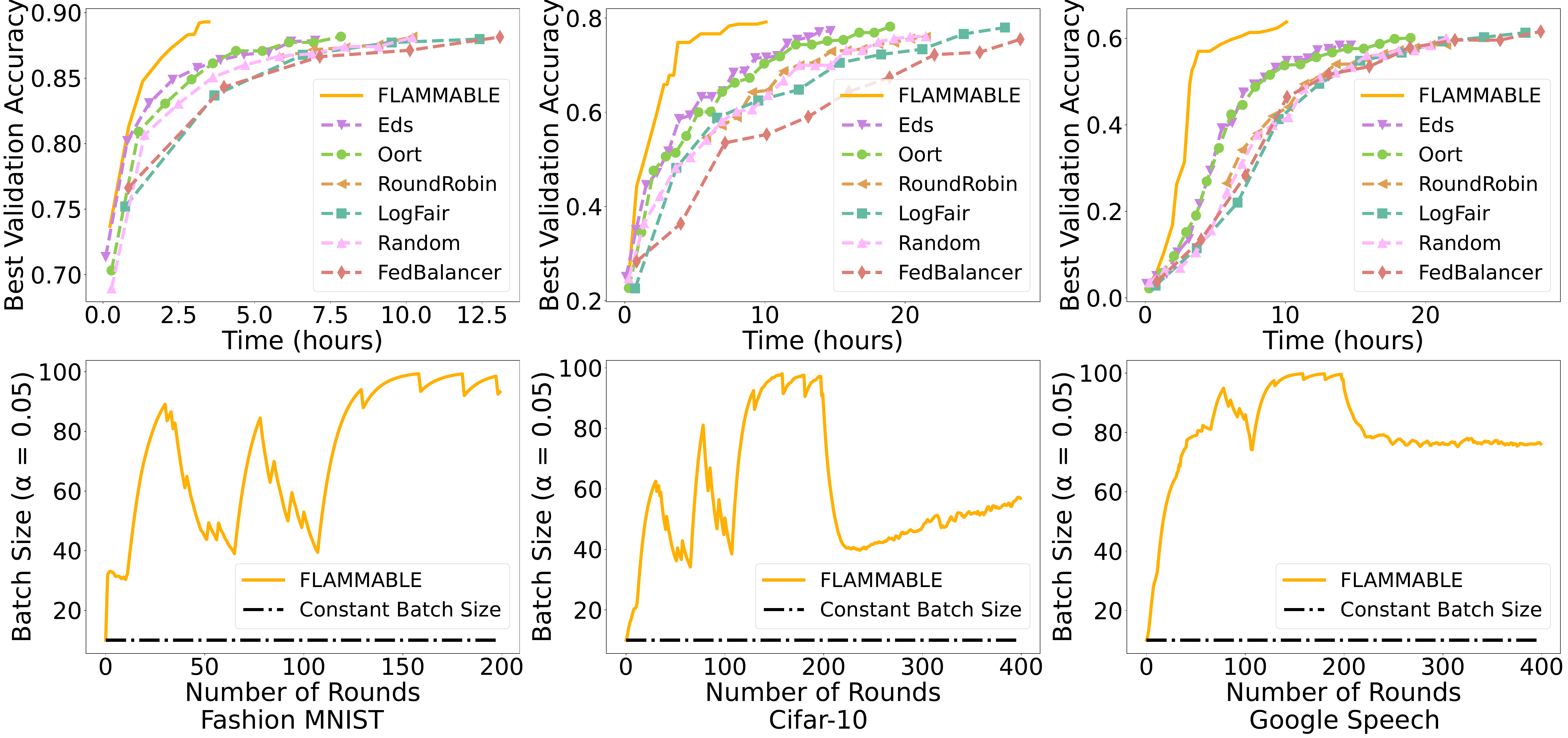}
  \caption{\name{} achieves the best time-to-accuracy performance (top). The average batch sizes chosen by \name{} for selected clients vary across datasets and rounds, smoothed using EMA with $\alpha=0.05$ (bottom).}
  \label{fig:group_a_all}
\end{figure*}

\name{} consistently shows high time-to-accuracy performance on all datasets, which we attribute to two features of our framework: batch size adaptation and multi-model engagement.

\textbf{Batch size adaptation.}
The bottom row of Figure~\ref{fig:group_a_all} shows the average batch size for each dataset in Group A chosen by \name{} during the training process.

\begin{figure*}[h]
    \centering
    \begin{subfigure}[t]{0.33\textwidth}
        \centering
        \includegraphics[width=\textwidth]{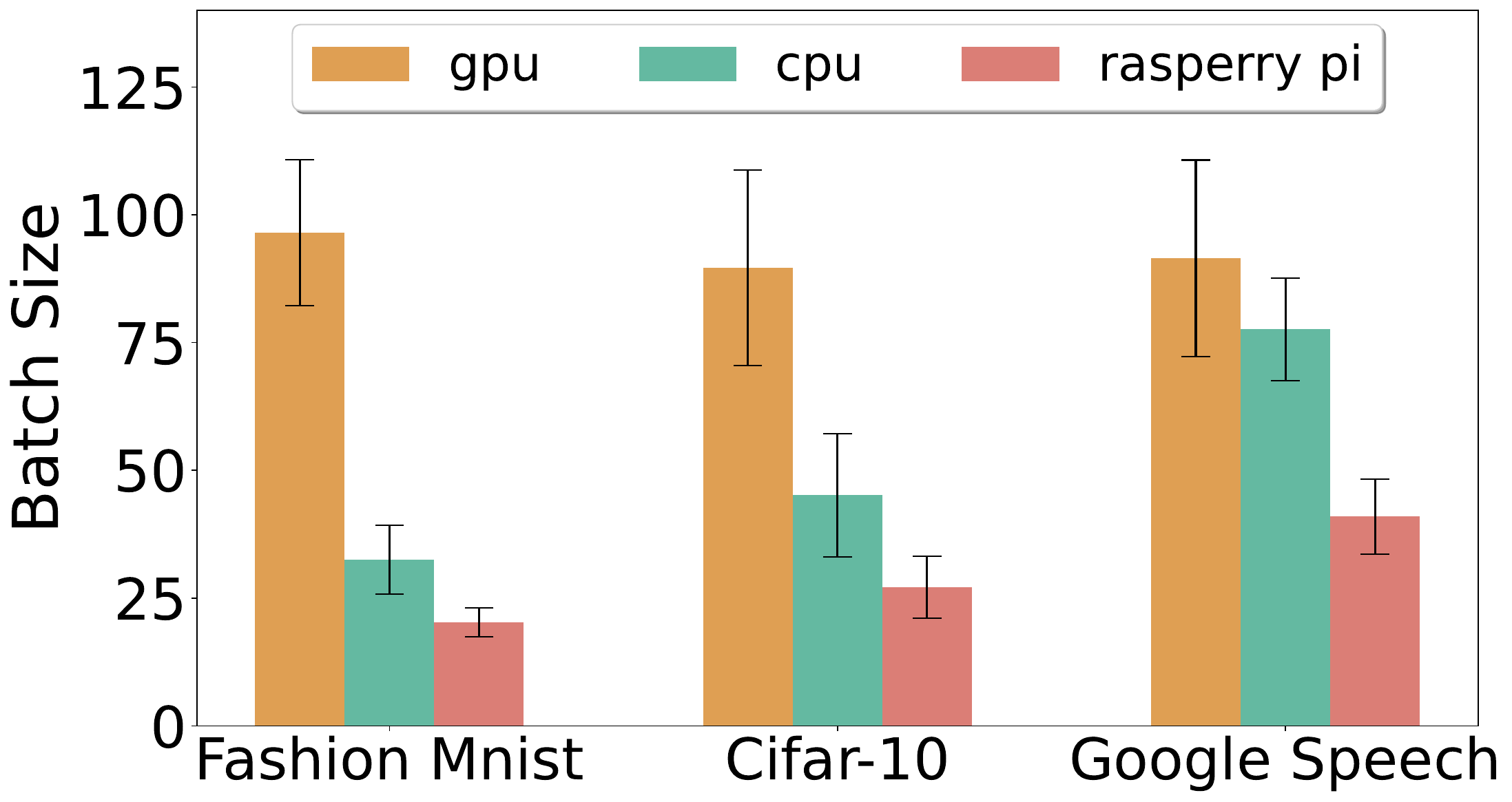}
        \caption{Group A}
        \label{fig:device_bsz:subfig1}
    \end{subfigure}
    \hfill
    \begin{subfigure}[t]{0.33\textwidth}
        \centering
        \includegraphics[width=\textwidth]{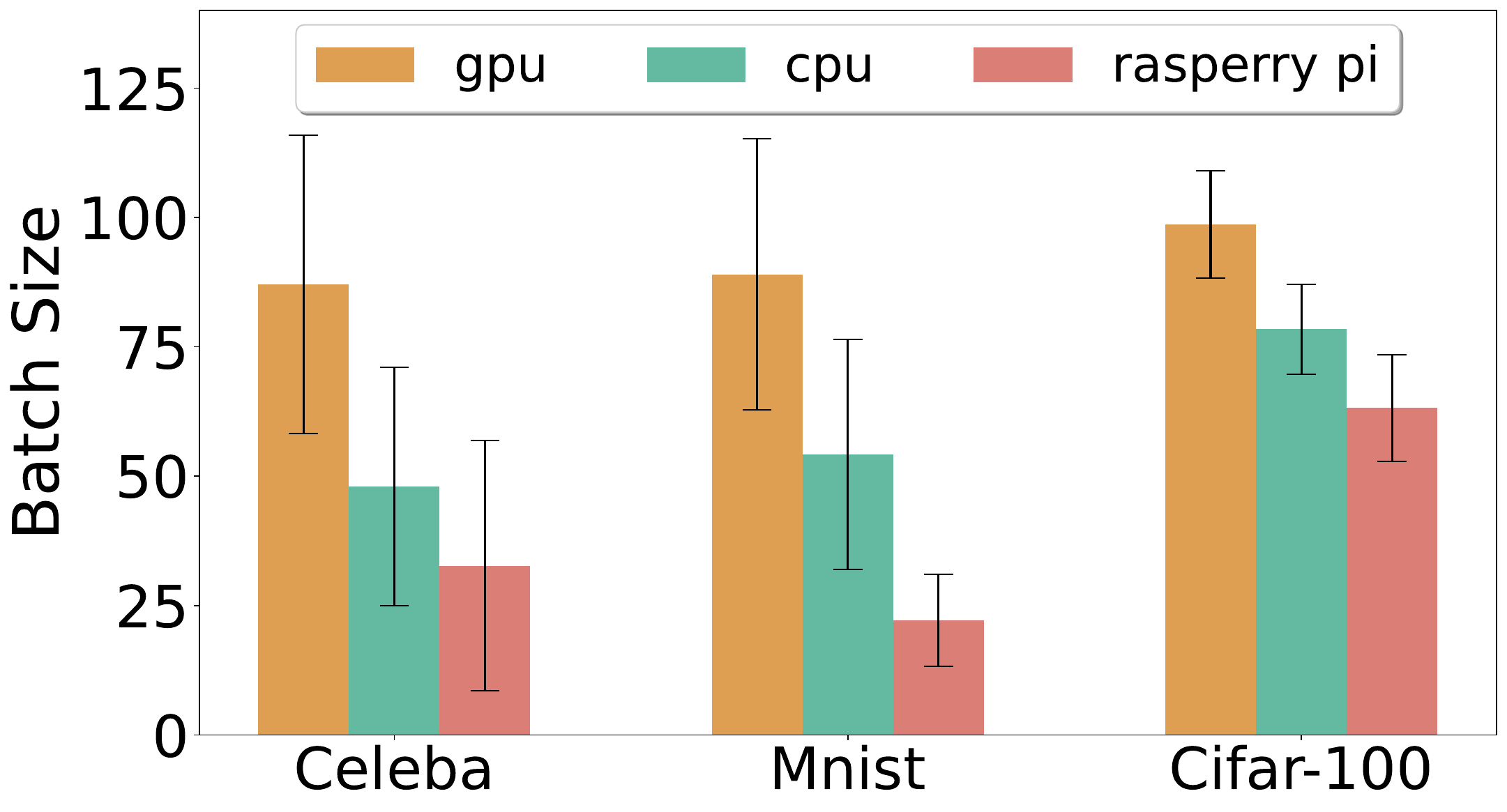}
        \caption{Group B}
        \label{fig:device_bsz:subfig2}
    \end{subfigure}
    \hfill
    \begin{subfigure}[t]{0.33\textwidth}
        \centering
        \includegraphics[width=\textwidth]{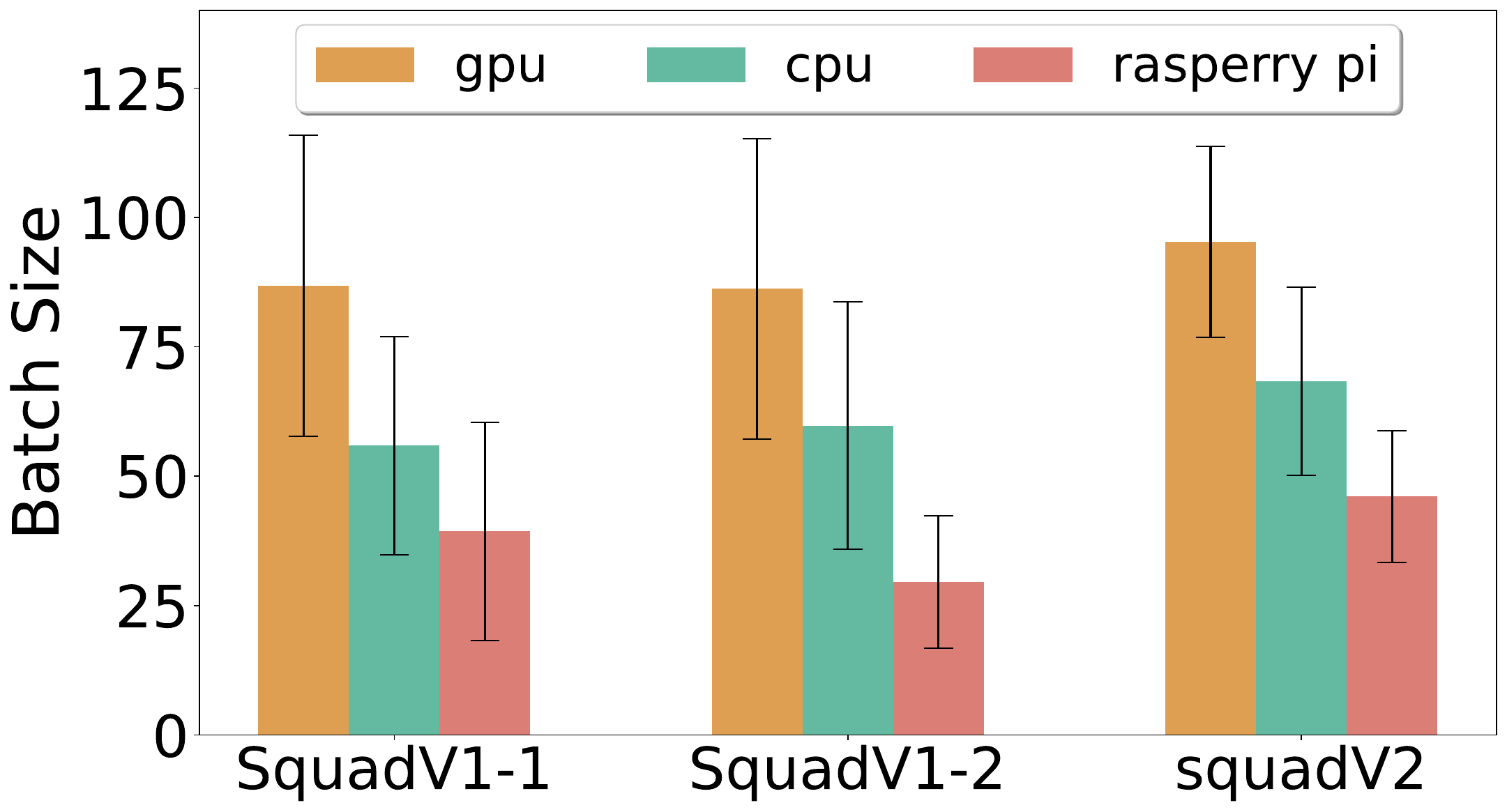}
        \caption{Group C}
        \label{fig:device_bsz:subfig3}
    \end{subfigure}
    \caption{The batch sizes (averaged across training rounds) chosen by \name{} vary with device types and datasets.}
    \label{fig:device_bsz}
\end{figure*}

We observe that the batch sizes vary over time. \name{} picks smaller batch sizes at the beginning and gradually increases the batch sizes. This choice is driven by the initially low gradient noise scale, where larger batch sizes could diminish statistical efficiency and even impinge on the final accuracy due to fewer model updates. The graph's jaggedness reflects \name{}'s calibration between clients' system throughput and data quality. As shown in Figure~\ref{fig:device_bsz}, clients with high system throughput benefit from larger batch sizes, while for clients with high data quality but limited compute capacity, optimal batch sizes tend to be smaller, as the benefits of increased throughput are outweighed by a decline in statistical efficiency. A notable decline in batch sizes for CIFAR-10 and Google Speech occurs when Fashion MNIST completes at round 200.
A fast client's utility is the sum of its utility for each model it can handle before the deadline. With fewer models, its utility may be surpassed by a slow client's high utility for just one model. As a result, \name{} favors these slower clients which typically have smaller batch sizes.


We further notice that batch sizes vary across clients and datasets. Figure~\ref{fig:device_bsz} compares the batch size chosen for clients of different device types on different datasets in Group A. We clearly see that, as we would intuitively expect, clients with more powerful device types (i.e., the CPU and GPU) choose larger batch sizes for the same dataset.
In addition, clients of the same device type adopt different batch sizes for different datasets.
%
This complexity confirms that manually fine-tuning batch sizes each round is impractical: the optimal batch sizes shift depending on the clients, datasets, and training phase, showing the benefits of our proposed automated batch size adaptation. 


\begin{figure*}[h]
    \centering
    \begin{subfigure}[t]{0.33\textwidth}
        \centering
        \includegraphics[width=\textwidth]{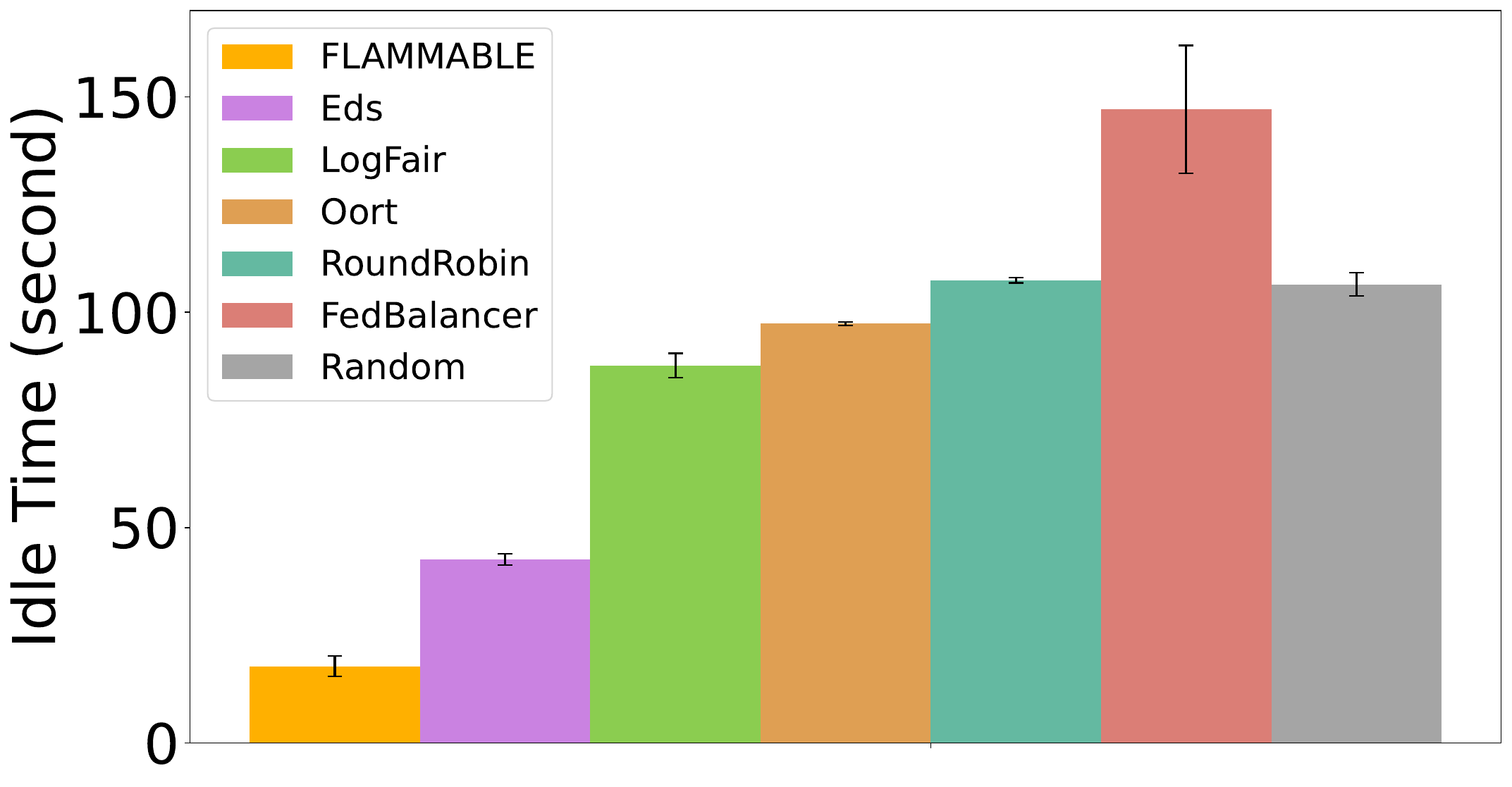}
        \caption{Group A}
        \label{fig:idle_percent:subfig1}
    \end{subfigure}
    \hfill
    \begin{subfigure}[t]{0.33\textwidth}
        \centering
        \includegraphics[width=\textwidth]{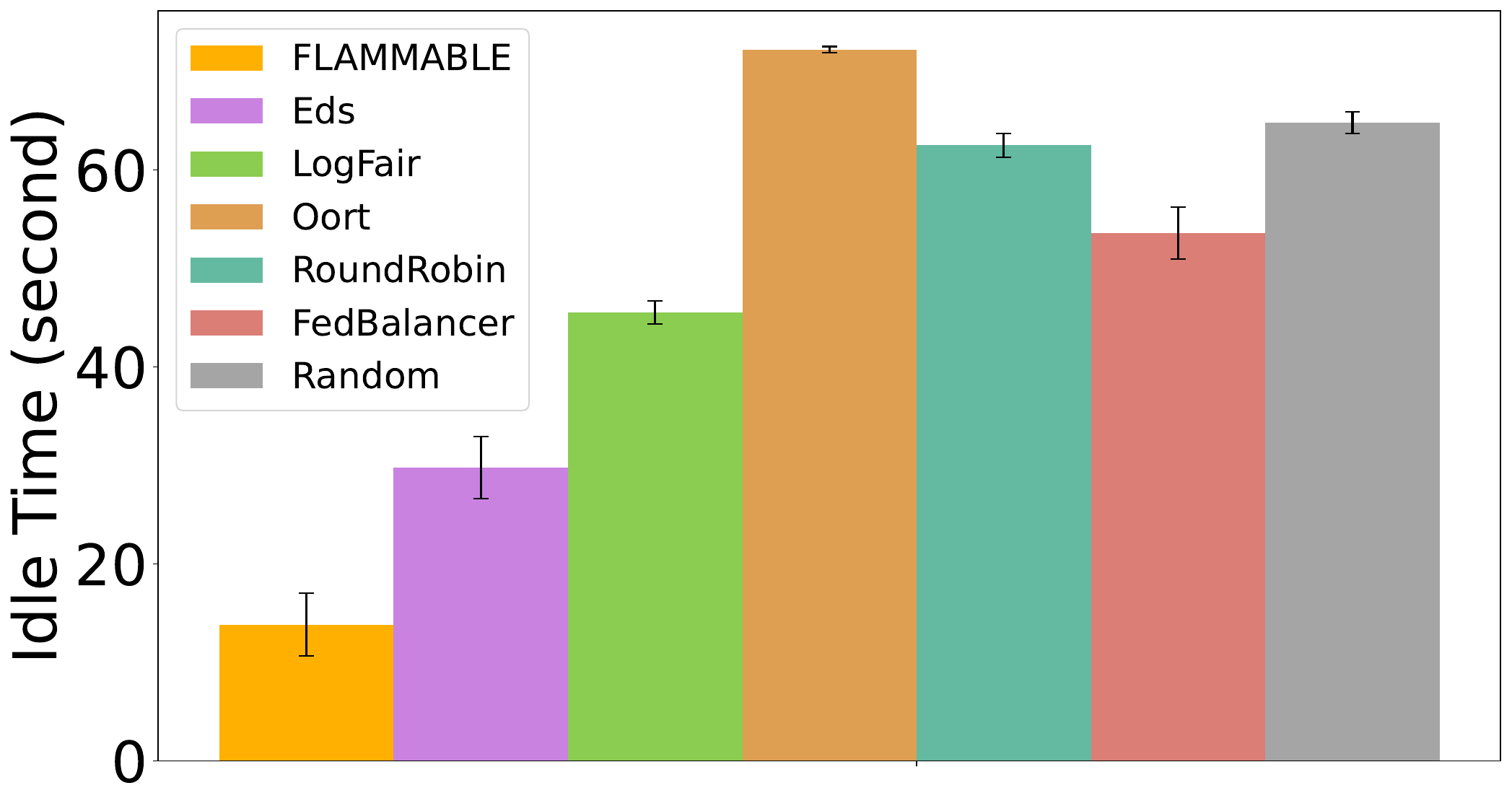}
        \caption{Group B}
        \label{fig:idle_percent:subfig2}
    \end{subfigure}
    \hfill
    \begin{subfigure}[t]{0.33\textwidth}
        \centering
        \includegraphics[width=\textwidth]{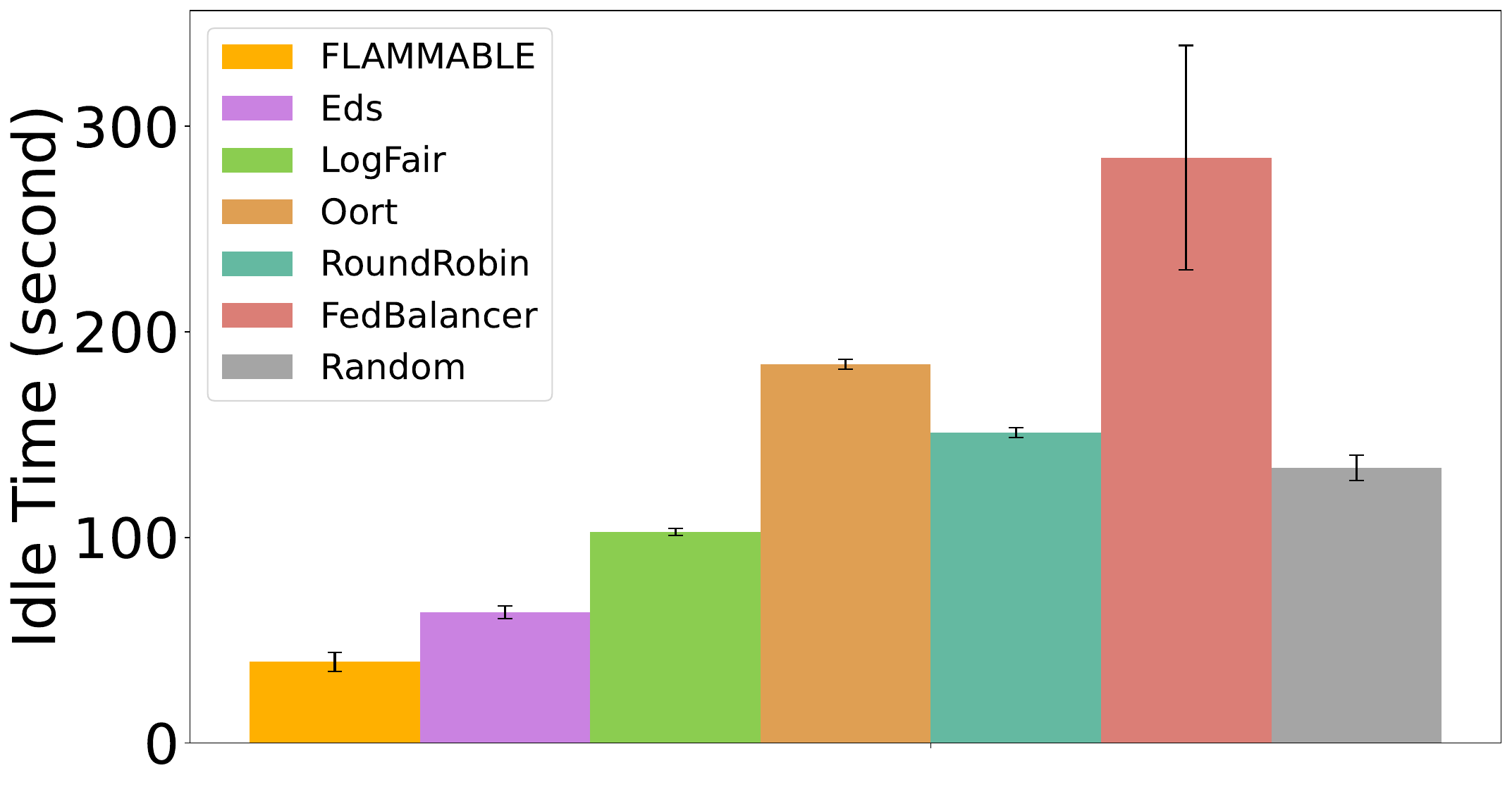}
        \caption{Group C}
        \label{fig:idle_percent:subfig3}
    \end{subfigure}
     \caption{\name{} can reduce client idle time (averaged across training rounds).}
  \label{fig:idle_percent}
\end{figure*}

\textbf{Multi-model engagement.}
Figure~\ref{fig:idle_percent} compares the average percentage of idle time observed among all selected clients during each round, when a minimum of two models are operational. By engaging clients with multiple models in reach round, \name{} reduces the idle time by 16\%-60\% compared to our baselines, thereby improving the time-to-accuracy performance and the final accuracy due to the increased volume and diversity of the training data used for each model.


\begin{figure*}[h!]
    \centering
    \begin{subfigure}[t]{0.33\textwidth}
        \centering
        \includegraphics[width=\textwidth]{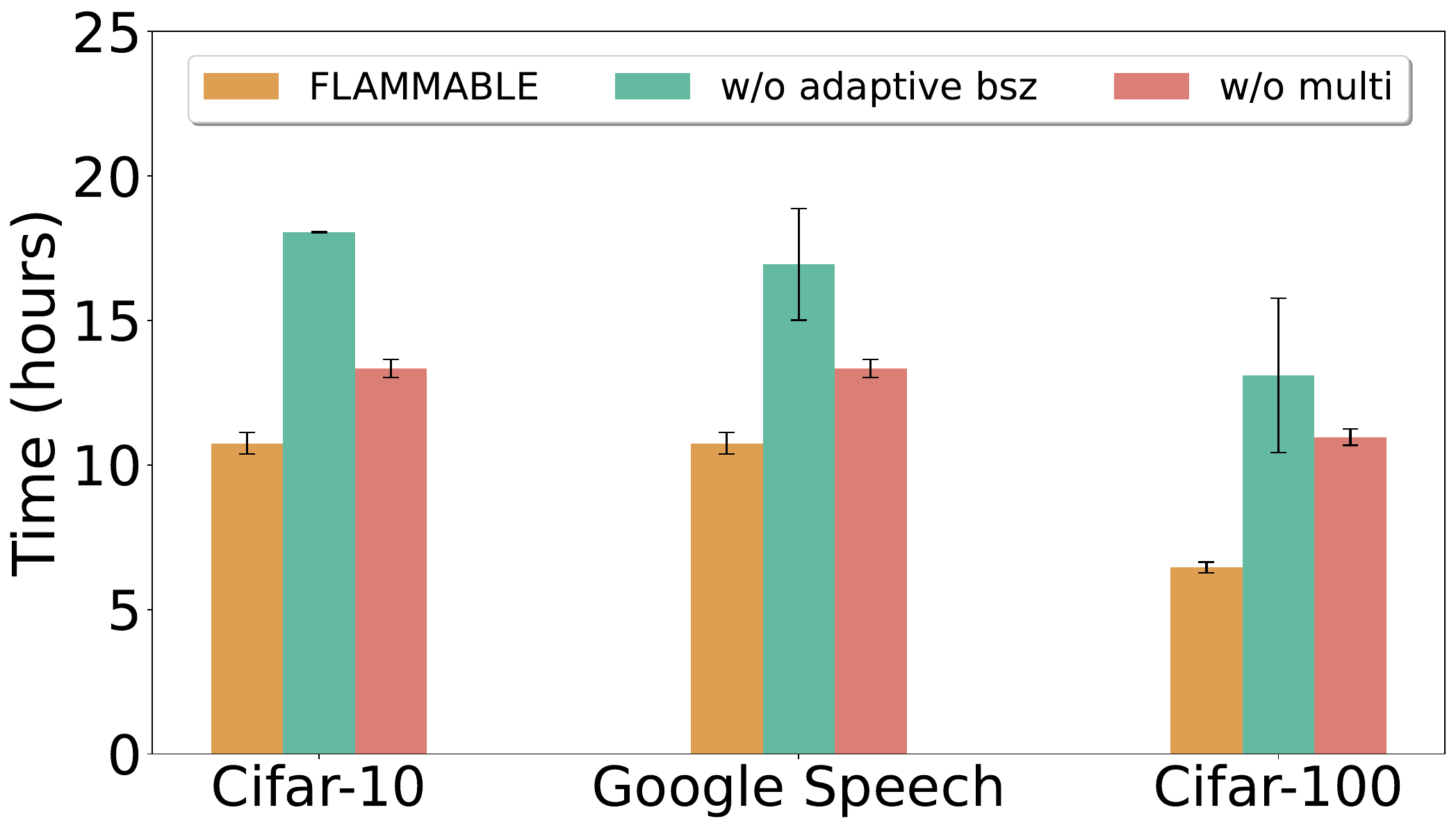}
        \caption{Group A}
        \label{fig:ablated_time:subfig1}
    \end{subfigure}
    \hfill
    \begin{subfigure}[t]{0.33\textwidth}
        \centering
        \includegraphics[width=\textwidth]{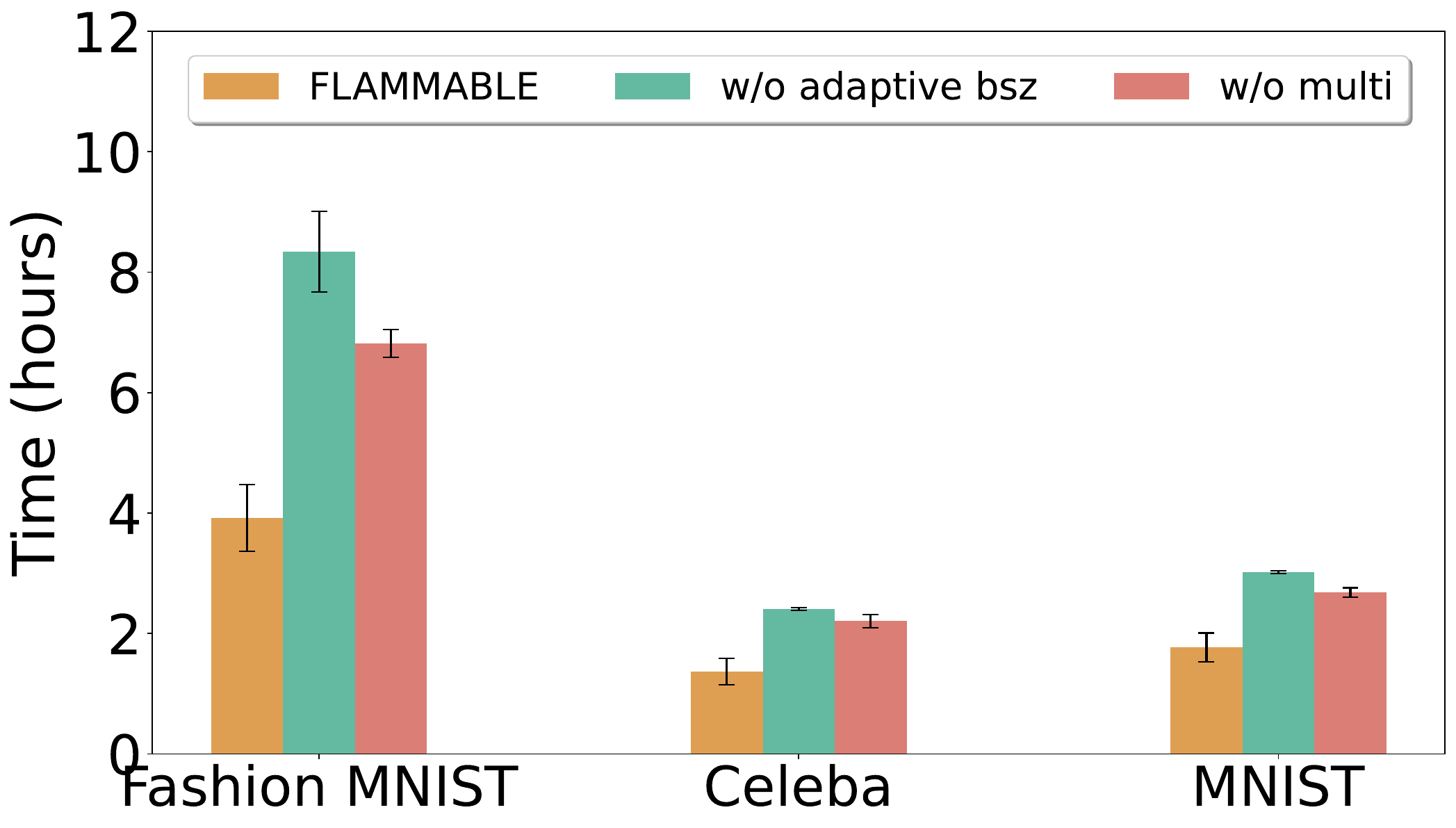}
        \caption{Group B}
        \label{fig:ablated_time:subfig2}
    \end{subfigure}
    \hfill
    \begin{subfigure}[t]{0.33\textwidth}
        \centering
        \includegraphics[width=\textwidth]{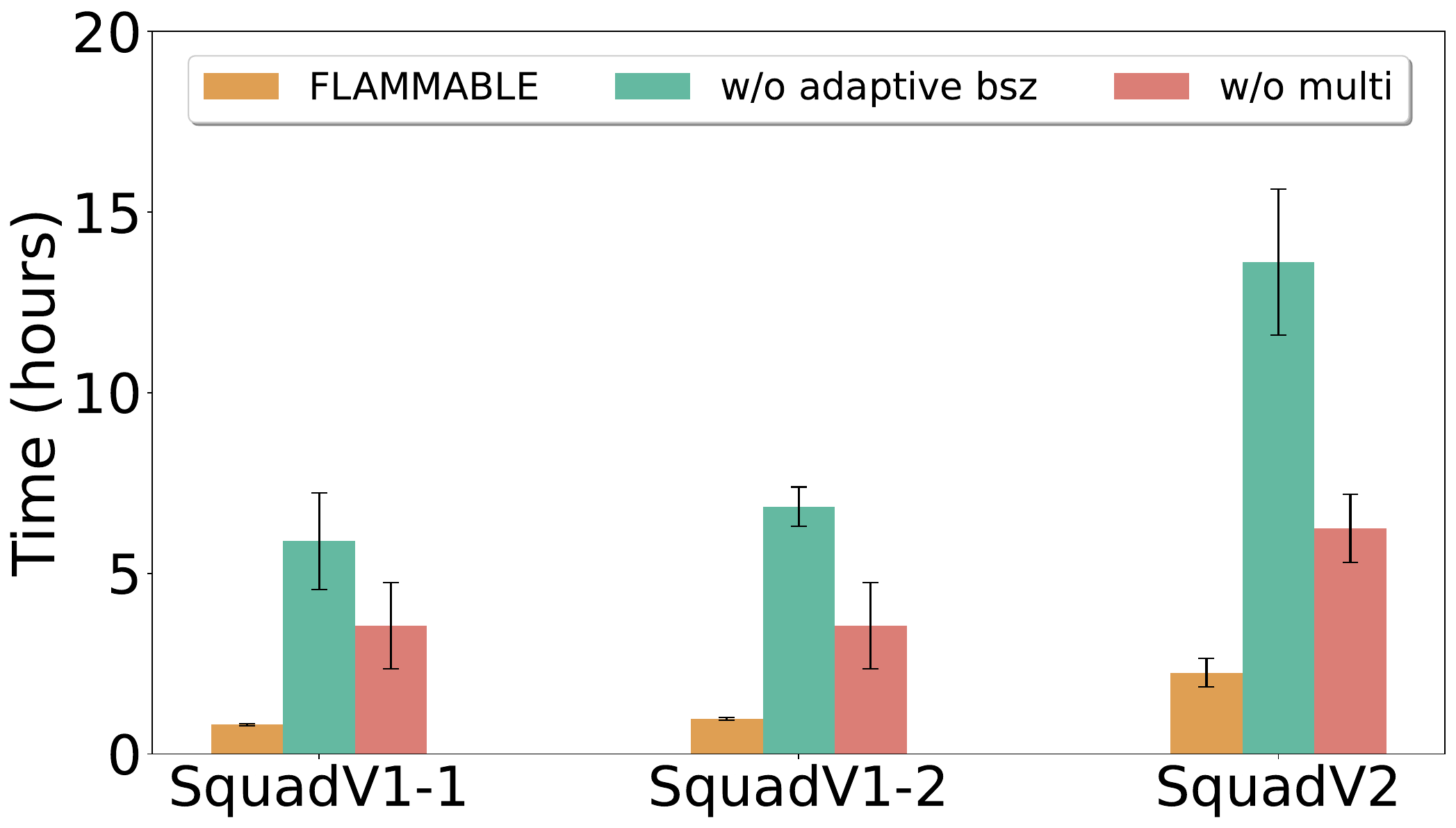}
        \caption{Group C}
        \label{fig:ablated_time:subfig3}
    \end{subfigure}
     \caption{Time-to-accuracy performance of \name{} without batch size adaptation or multi-model engagement.}
  \label{fig:ablated_time}
\end{figure*}

\textbf{Ablation study.} We show the importance of directly adopting each key strategy (batch size adaptation and multi-model engagement) by evaluating scenarios where only one is present. As shown in Figure~\ref{fig:ablated_time}, excluding batch size adaptation and multi-model engagement results in an average slowdown of 44\% and 32\% respectively, which underscores the efficacy of both strategies in optimizing performance.


\begin{figure}[h]
  \centering
  \includegraphics[width={\linewidth}]{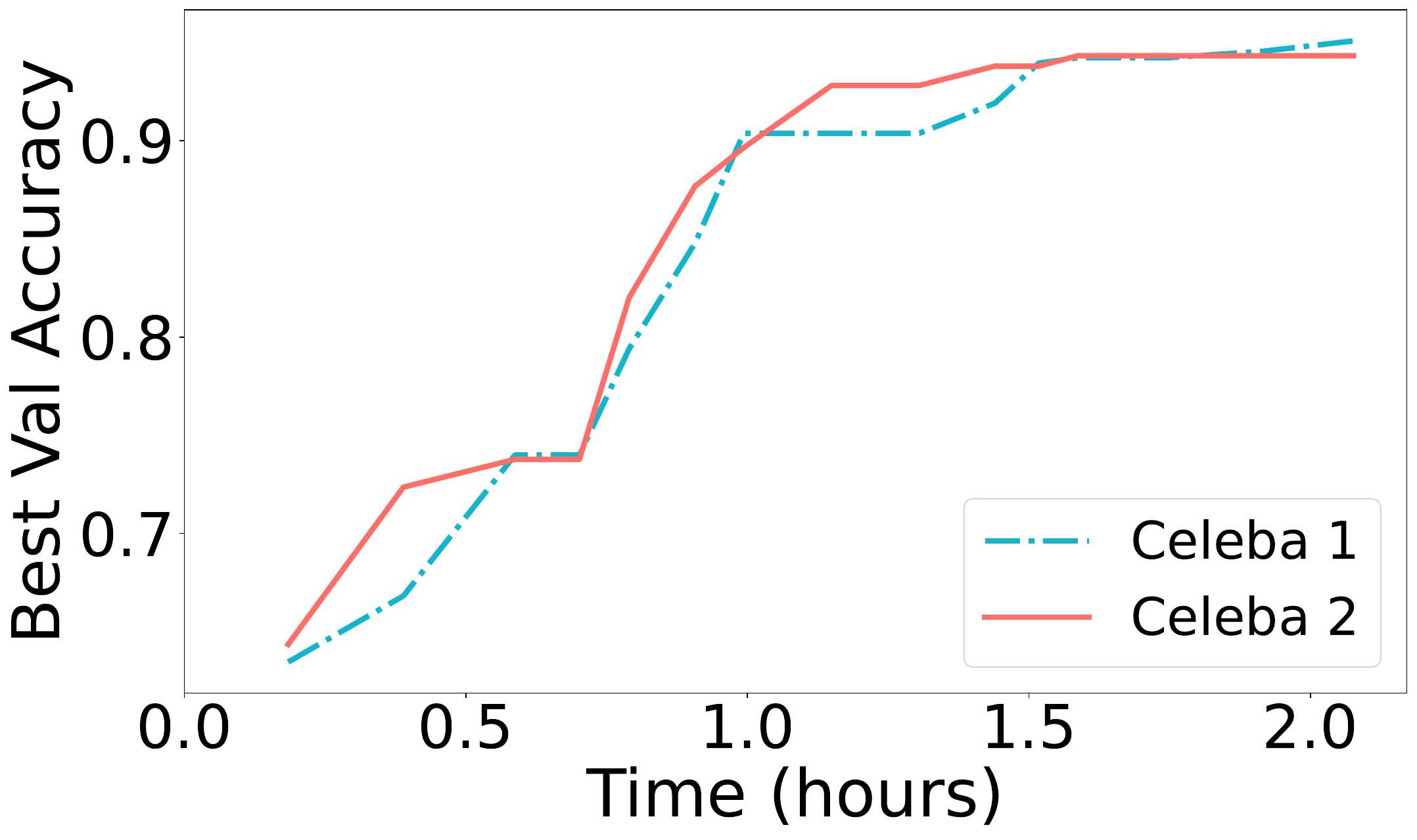}
  \caption{\name{} ensures fairness across models.}
  \label{fig:single_fair_celeba}
\end{figure}

\textbf{Model fairness.}
To assess the fairness of \name{}'s client selection policy across different models, we examine a scenario featuring two identical models. Ideally, both models should receive an equal distribution of clients, resulting in comparable time-to-accuracy performance. As depicted in Figure~\ref{fig:single_fair_celeba}, the performance curves of both models closely align, demonstrating that \name{} fairly allocates clients to them.

\subsection{Sensitivity Analysis}\label{eval:sensitivity}

\textbf{Number of participants.}
We assess \name{} with varying number of clients participating per round. Given EDS's consistent top performance among baselines across all models, we limit our comparison to \name{} versus EDS to conserve simulation resources. Table~\ref{tab:num_participant} reveals that \name{} consistently outperforms EDS in time-to-accuracy, with the performance gap widening as participant numbers grow. This trend stems from \name{}'s ability to allocate more clients to each job with increased participation, thereby accelerating the training process.

\begin{table}[h]
    \centering
    \scalebox{1.0}{
        \begin{tabular}{c|c|c|c}
            \hline
            \specialcell{Num.\\Participants} & \specialcell{Fashion MNIST} & Cifar-10 & \specialcell{Google Speech}\\
            \hline
             15 & 1.94$\times$ & 1.23$\times$ & 1.10$\times$ \\
             30 & 2.22$\times$ & 1.47$\times$ & 1.47$\times$ \\
             60 & 2.55$\times$ & 1.57$\times$ & 1.57$\times$ \\

            \hline
            \specialcell{Num.\\Participants} & \specialcell{Celeba} & Mnist & \specialcell{Cifar-100}\\
            \hline
             15 & 1.50$\times$ & 1.97$\times$ & 1.09$\times$ \\
             30 & 1.60$\times$ & 1.47$\times$ & 1.57$\times$ \\
             60 & 1.54$\times$ & 1.45$\times$ & 1.57$\times$ \\
            \hline
        \end{tabular}
    }
    \caption{Time-to-accuracy improvement of \name{} over EDS increases with a higher number of participants.}
    \label{tab:num_participant}
\end{table}


\textbf{Impact of uncertainty factor $\alpha$.}
In \name{}, the uncertainty factor $\alpha$ influences client selection, prioritizing those not recently chosen. As highlighted in Table~\ref{tab:uncertainty}, a smaller $\alpha$ value generally speeds up the training process at the cost of final accuracy. As training progresses, differences in client data quality (loss) diminish, and system throughput becomes the dominant factor in client utility. A smaller $\alpha$ favors faster clients, speeding up training but potentially decreasing final accuracy. To balance convergence time and model quality, we choose $\alpha=1$.

\begin{table}[h]
    \centering
    \scalebox{1.0}{
        
    \begin{tabular}{c|cc|cc|cc}
        \hline
        Dataset & \multicolumn{2}{c|}{Celeba} & \multicolumn{2}{c|}{Mnist} & \multicolumn{2}{c}{Cifar-100}\\
    \hline
             $\alpha$&  Time &  Acc. &  Time&  Acc.&  Time& Acc.\\
             0.1 &  0.58&  94.7&  0.83&  94.0&  8.47& 52.1\\
             \textbf{1} &  0.68&  95.0&  1.33&  95.2&  6.21& 55.6\\
             10 &  1.1&  95.0&  1.91&  95.3&  11.4& 56.0\\
             
    \hline
    Dataset & \multicolumn{2}{c|}{Fashion Mnist} & \multicolumn{2}{c|}{Cifar-10} & \multicolumn{2}{c}{Google Speech}\\
    \hline
             $\alpha$&  Time &  Acc. &  Time&  Acc.&  Time& Acc.\\
             0.1 &  1.8 &  88.5&  5.9&  75.4&  5.9 & 65\\
             \textbf{1} &  3.9&  89.3&  10.98&  79.8&  10.98& 65.3\\
             10 &  5.63&  89.2&  13.54&  80.9&  13.54& 65.9\\
    \hline
    
    \end{tabular}
    }

    \caption{Impacts of uncertainty factor on \name{}: smaller $\alpha$ reduces runtime (hours) but degrades final accuracy (\%).}
    \label{tab:uncertainty}
\end{table}



\section{Conclusion}
\label{sec:conclusion}

In this work, we propose FLAMMABLE, an MMFL framework that optimizes the time-to-accuracy performance for parallel model training. Unlike existing MMFL solutions, FLAMMABLE (i) intelligently adapts client batch sizes in each training round to expedite the training progress; and (ii) enables multi-model engagement per client to minimize client idle time and enhance the overall time-to-accuracy. To facilitate evaluation, we implement a custom benchmarking platform, which is the first platform to fully support MMFL scenarios to the best of our knowledge. Through evaluations across diverse datasets, FLAMMABLE demonstrates an improvement in the MMFL time-to-accuracy performance by 1.1$\sim$5.2$\times$, while enhancing the final model accuracy by 1.3$\sim$5.4\%. These results outperform both na\"ive extensions of single-model FL algorithms and prior MMFL approaches.

Future works on \name{} include co-optimization of other hyperparameters such as model learning rates, which complement our work in this paper and can be easily incorporated into our evaluation platform. We additionally plan to leverage the benchmarking platform to evaluate \name{}'s performance on other models and in more realistic settings such as dynamic model arrival; while these settings can easily be accommodated in \name{}, empirically evaluating their performance may be of interest.

\newpage
\newpage
\bibliographystyle{ACM-Reference-Format}
\bibliography{section/main}

\end{document}